\documentstyle[pre,aps,graphics,epsfig]{revtex}

\begin{document}
\title{{\bf Directed Percolation has Colors and Flavors}}
\author{{\sc Hans-Karl\ Janssen}}
\address{Institut f\"{u}r Theoretische Physik III, Heinrich-Heine-Universit\"{a}t,\\
40225 D\"{u}sseldorf, Germany }
\date{\today}
\maketitle

\begin{abstract}
A model of directed percolation processes with colors and flavors
that is equivalent to a population model with many species near
their extinction thresholds is presented. We use renormalized
field theory and demonstrate that all renormalizations needed for
the calculation of the universal scaling behavior near the
multicritical point can be gained from the one-species Gribov
process (Reggeon field theory). In addition this universal model
shows an instability that generically leads to a total asymmetry
between each pair of species of a cooperative society, and finally
to unidirectionality of the interspecies couplings. It is shown
that in general the universal multicritical properties of
unidirectionally coupled directed percolation processes with
linear coupling can also be described by the model. Consequently
the crossover exponent describing the scaling of the linear
coupling parameters is given by $\Phi =1$ to all orders of the
perturbation expansion. As an example of unidirectionally coupled
directed percolation, we discuss the population dynamics of the
tournaments of three colors.

\end{abstract}



\section{Introduction}

Nonequilibrium processes, their stationary states and their phase
transitions have been of considerable interest in na\-tu\-ral science as
well as in medicine and sociology for many years. Here we are interested in
processes that can be modelled by growth and decay of populations with
spatially local interaction rules. The transition between survival and
extinction of a population is a nonequilibrium continuous phase transition
phenomenon and is characterized by universal scaling laws. It is well known
that also for systems far from equilibrium the concept of universality
classes with respect to their critical properties in the vicinity of a
continuous phase transition is applicable. For the description of
transitions in systems that show active and absorbing inactive states,
percolation models play an outstanding role. Some years ago it was
conjectured \cite{Ja81,Gr82} that Markovian growth mo\-dels with
one-component order parameters displaying a transition into an absorbing
state in the absence of any special conservation law generically belong to
the universality class of directed percolation (DP). Besides DP \cite
{BrHa57,CaSu80,Ob80} this universality class includes e.g. Reggeon field
theory (RFT) \cite{Gribov,Mo78,GrassbGr}, the contact process \cite
{Ha74,Li85,JeDi94}, certain cellular automata \cite{Ki83} and some catalysis
models \cite{ZGB86,GLB89} (for a recent review of DP processes see \cite
{Hin00}).

Despite the fact that a large variety of different models belong to the DP
universality class, {\em there is still no experiment where the critical
behavior of DP was seen} \cite{Gra96}. In a recent paper \cite{Hin99},
Hinrichsen compares suggested experiments and discusses possible reasons why
the observation of DP critical exponents is obscured or even impossible. One
of these reasons might be that the basic feature of the DP class, the
existence of an absorbing state, is quite difficult to realize in nature.
Small fluctuations will always affect this state and may be strong enough to
soften the transition like a small particle source, which works as an
external field \cite{JKO98}. Another reason might be the influence of
spatial quenched disorder which is abundant in reality. We have shown \cite
{Ja97b} that in contrast to equilibrium systems, the critical scaling
properties of DP processes are not only altered by frozen randomness, but
fully destroyed.

For the analytic description of universal behavior near a cri\-ti\-cal
nonequilibrium transition, it is often useful to model the universality
class by mesoscopic stochastic processes involving the order parameter and
other relevant fields. In case of the DP class a representation by the
Langevin equation for the time-development of the particle density, the
Gribov process (the stochastic version of the so-called Schl\"{o}gl model
\cite{Sch72}), is appropriate. The name Gribov process was coined by
Grassberger who showed that RFT is a Markov process in disguise rather
than a quantum theory \cite{GrassbGr}. On the level of a formulation of
stochastic processes by means of path integrals, there is superficially no
difference between RFT and the Gribov process. However RFT uses creation
and annihilation operators for particles as the principal fields in
contrast to the particle density and its conjugate response field in the
Gribov process. Microscopically the RFT-description of DP starts with
special reactions between diffusing individuals on a given $d$-dimensional
lattice such as birth: $X\rightarrow X+X$, competition: $X+X\rightarrow
X$, and death: $ X\rightarrow 0$. These reactions are represented by a
master equation that is mapped onto a second-quantized bosonic operator
representation, which is in turn mapped onto a bosonic field theory using
the continuum limit \cite {EVForm}. At the critical point the rates of
birth and death have to balance to yield a vanishing overall production of
individuals. But this condition leads to strong local correlations and the
microstates of the system consist typically of clusters of individuals
embedded in the vacuum \cite{Gr82}. Thus, a fluctuating density
description for the Gribov process is appropriate on a mesoscopic level.
Note that {\em the replacement of the spontaneous single particle death
reaction by a two particle reaction $X+X\rightarrow 0$ leads to strong
anticorrelations because now the birth rate itself has to vanish at the
critical point}. In this case, the microstates consist of separated lonely
wanderers that only sporadically interact and no clustering of individuals
sets in. {\em For such problems of branching and annihilating random walks
one is forced to use the creation-annihilation operator formulation}
\cite{CaTa98} and a naive stochastic density description for the
fundamental field would lead to wrong results. As a general rule: one can
show that branching processes lead to positive correlations and
annihilation processes to anticorrelations with the exception of
spontaneous decay, which does not generate any correlations. Thus, {\em
nonvanishing branching and spontaneous decay are needed to yield positive
correlations together with the possibility of a vanishing overall
production of particles at the critical point. In such cases a mesoscopic
density description is correct.}

Instead of considering only one species of particles as is usually done for
processes belonging to the DP class, it is of interest to introduce
processes with several interacting species as for instance in mathematical
biology \cite{HoSi88,Mu89}, which are also of relevance for a special model
of surface growth \cite{AEHM96}. It is the purpose of this paper to describe
the detailed field-theoretic investigation based on renormalization group
methods of such {\em colored and flavored directed percolation processes}
(MDP) realized by the Gribov process for several species. We group different
colored species in the same flavor class, if they have equal transport
properties.

In the next chapter we introduce the model and define its renormalization
and the one-loop calculation in the third chapter. In chapter IV, we
present the general renormalization group analysis and find the asymptotic
scaling behavior to one-loop order. In chapter V, we include the two-loop
results of the appendices and show the crossover to unidirectionality of
the couplings between the different species. In chapter VI, we present our
considerations on symmetries and the general fixed point properties of the
model. We show that the permutation symmetry of a multicolored process is
spontaneously broken. A brief account of this work has been presented in
\cite{Ja97}. In addition, in chapter VII, we will show that the universal
multicritical features of a recently introduced model of unidirectional
coupled directed percolation processes (UCDP) by T\"{a}uber et al. \cite
{THH98}, which contains an additional linear coupling between the species,
is completely described by the MDP class. In the last chapter we discuss
the results and give an application to biomathematics. Three appendices
present technical details, e.g.\ the $\varepsilon $-expansion of the
DP-exponents, known for a long time, but as yet unpublished.

\section{The model}

The mesoscopic description of the dynamics of physical systems is based on
a correct choice of the complete set of fundamental slowly-developing
fields. In general these are the order parameter densities and the
densities of conserved quantities. The multispecies processes under
consideration are completely described by the particle densities
$n\left({\bf x},t\right) =\left( n_{1}\left({\bf
x},t\right),n_{2}\left({\bf x},t\right),\ldots \right)$ of the percolating
colored and flavored individuals. We assume that there does not exist any
conservation law.

Next one has to find out the general form of the stochastic equations of
motions of the fundamental fields as timelocal (Markovian) Langevin
equations. These Langevin equations have to respect symmetries and general
principles characterizing the universality class under consideration. The
MDP-class is characterized by the following four principles:

\begin{enumerate}
\item  Errorfree self-reproduction (``birth'') and spontaneous annihilation
(``death'') of individuals. The rates for birth and death may be different
for each color.

\item  Interaction between the individuals (``competition'', ``saturation'')
       with color-dependent couplings.

\item  Diffusion (``motion'', ``spreading'') of the individuals in a $d$-dimensional
       space with flavor-depending transport coefficients.

\item  The states with at least one extinct color are absorbing.
\end{enumerate}

In the language of chemistry, the MDP may be realized microscopically by an
autocatalytic reaction scheme of the form $X_{\alpha }\leftrightarrow
2X_{\alpha }$, $X_{\alpha }\rightarrow 0$, $X_{\alpha }+X_{\beta
}\rightarrow kX_{\alpha }+lX_{\beta }$, where the last reaction subsumes the
interactions of individuals with colors $\alpha $, $\beta ,$ and $k,l$ may
be the integers $0,1$. A description in terms of particle densities
typically arises from a coarse-graining procedure where a large number of
microscopic degrees of freedom are averaged out. The influence of these is
simply modelled by Gaussian noise-terms in the Langevin equation which
however have to respect the absorbing state condition. The stochastic
reaction-diffusion equations for the particle densities in accordance with
the four principles given above are of the form
\begin{equation}
\partial _{t}n_{a}({\bf x},t)=\lambda _{\alpha }\nabla ^{2}n_{\alpha }({\bf x
},t)+R_{\alpha }(n({\bf x},t))\,n_{\alpha }({\bf x},t)+\zeta _{\alpha
}({\bf x},t)\ ,  \label{1}
\end{equation}
where the first term on the right hand side models the (diffusive) motion,
and the $R_{\alpha }$ are the overall reproduction rates of the particles
with color $\alpha $. These deterministic terms are constructed proportional
to $n_{\alpha }$ in order to ensure the existence of an absorbing state for
each species. Near the absorbing transition the particle densities $n$ are
small quantities. Expanding the rates $R_{\alpha }$ in powers of $n$ results
in
\begin{equation}
R_{\alpha }(n)=-\lambda _{\alpha }\biggl(\tau _{\alpha }+
\frac{1}{2}\sum_{\beta }g_{\alpha \beta }\ n_{\beta }{\bf +}\cdots
\biggr). \label{2}
\end{equation}
The Gaussian noises $\zeta _{\alpha }\left( {\bf r},t\right) $ must also
respect the absorbing state condition, whence
\begin{eqnarray}
\langle \zeta _{\alpha }({\bf x},t)\zeta _{\beta }({\bf x}^{\prime
},t^{\prime })\rangle &=&2D_{\alpha \beta }(n({\bf x},t))\delta ({\bf x}-
{\bf x}^{\prime })\delta (t-t^{\prime })+\cdots  \nonumber \\ &=&\lambda
_{\alpha }g_{\alpha }\delta _{\alpha ,\beta }\,n_{\alpha }({\bf x}
,t)\delta ({\bf x}-{\bf x}^{\prime })\delta (t-t^{\prime })+\cdots \ .
\label{3}
\end{eqnarray}
Subleading terms in the expansions (\ref{2},\ref{3}) as well as additional
terms with derivatives of the spatial $\delta $-function in the first line
of Eq.\ (\ref{3}) are not displayed. It can be shown that they are
irrelevant in the renormalization group sense as long as the stability
condition $\sum_{\alpha ,\beta }\lambda _{\alpha }g_{\alpha \beta
}n_{\alpha }n_{\beta }\geq 0$ for all $n_{\alpha }\geq 0$ is fulfilled.
The breakdown of this condition signals the occurrence of a discontinuous
transition to compact growth and the appearance of tricritical phenomena
at the border between first and second order transitions. Such a behavior
is expected in microscopic models based on more complicated particle
reactions \cite{OhKe87}. The ``temperature'' variables $\tau _{\alpha }$
measure the difference of the rates of death and birth of the color
$\alpha $. Thus the temperatures may be positive ore negative. We are
interested in the case where all $\tau _{\alpha }\approx 0$ (up to
fluctuation corrections) which defines the multicritical region. Under
these conditions all the species live on the border of extinction.

The next step is a mean field investigation of homogeneous steady state
solutions of the equations of motion. Neglecting all fluctuations in Eq.\
(\ref{1}) and using the expansion (\ref{2}) up to first order in $n,$ we
find easily that all $M_{\alpha }:=\langle n_{\alpha }({\bf x},t)\rangle
_{steady\ state}=0$ as long as all $\tau _{\alpha }>0$, meaning the vacuum
is absorbing for each color. As soon as some of the temperatures become
negative, stationary solutions with $M_{\alpha }>0$ emerge, satisfying the
corresponding equations $\sum_{\beta }\,g_{\alpha \beta }M_{\beta }=-2\tau
_{\alpha }$. Thus, in general a multitude of hypersurfaces of first and
second order transitions exist in the phase space spanned by the relevant
temperature variables $\left\{ \tau _{\alpha }\right\} ,$ which separate
the phases where a specific color becomes extinct. Whenever $(\sum_{\beta
}\,g_{\alpha \beta }M_{\beta }+2\tau _{\alpha })$ changes from a negative
to a positive value, the order parameter $M_{\alpha }$ undergoes a
continuous or a discontinuous phase transition from an inactive absorbed
state with $M_{\alpha }=0$ to an active state with $M_{\alpha }>0$. All
hypersurfaces of phase transitions meet in the multicritical point where
all temperatures are zero. All homogeneous states are globally stable
because the evolution of the total particle density of homogeneous states
in time is given by $d(\sum_{\alpha }M_{\alpha })/dt=-\sum_{\alpha
}\lambda _{\alpha }\tau _{\alpha }M_{\alpha }-\frac{1}{2}\sum_{\alpha
,\beta }\lambda _{\alpha }g_{\alpha \beta }M_{\alpha }M_{\beta }$. Thus
all solutions of the mean field equations of motion are bounded to a
finite region in the space of positive $M_{\alpha }$ as long as the
stability condition mentioned in the foregoing paragraph holds.

In the following we focus on the effect of fluctuations on the scaling
behavior of correlation and response functions in the vicinity of the
multicritical point where the strongly relevant parameters \{$\tau
_{\alpha } $\} are small. In order to apply field-theoretic methods and
the renormalization group equation in conjunction with an $\varepsilon
$-expansion about the upper critical dimension \cite
{Am84,ZiJu93,BJW76,DDP78,Ja79}, it is convenient to use the path-integral
representation of the underlying stochastic processes $n({\bf x}
,t)=\{n_{\alpha }({\bf x},t)\}$ \cite{Ja79,DeDo76,Ja76,Ja92}. With the
imaginary-valued response fields denoted by $\widetilde{n}({\bf x},t)=\{{
\widetilde{n}}_{\alpha }({\bf x},t)\}$, the generating functional of the
connected response and correlation functions, the Greens functions, takes
the form
\begin{equation}
{\cal W}\bigl[h,\widetilde{h}\bigr]=\ln \int {\cal D}\bigl[\widetilde{n},n
\bigr]\exp \biggl(-{\cal J}\bigl[\widetilde{n},n\bigr]+\int d^{d}x\int dt
\bigl(hn+\widetilde{h}\widetilde{n}\bigr)\biggr)\ .  \label{PathInt}
\end{equation}
The response fields $\widetilde{n}\left( {\bf x},t\right) $ correspond to
the conjugated au\-xi\-li\-ary variables of the operator formulation of
statistical dynamics by Kawasaki \cite{Kaw71} and Martin, Siggia, Rose
\cite {MSR72}). The dynamic functional ${\cal
J}\bigl[\widetilde{n},n\bigr]$ and the functional measure ${\cal
D}\bigl[\widetilde{n},n\bigr]$, which in symbolic notation is proportional
to $\prod_{{\bf x},t}\bigl(d\widetilde{n}( {\bf x},t)dn({\bf x},t)\bigr)$,
is understood to be defined using a prepoint (Ito) discretization with
respect to time. The prepoint discretization leads to the causality rule
$\theta (t\leq 0)=0$ in response functions which forbids response
propagator loops in the diagrammatical perturbation expansion
\cite{Ja79,Ja92}.

The Langevin equations (\ref{1}-\ref{3}) are recast as a dynamic functional

\begin{eqnarray}
{\cal J} &=&\int dtd^{d}x\biggl(\sum_{\alpha }\widetilde{n}_{\alpha
}\Bigl(\partial _{t}-\lambda _{\alpha }\nabla ^{2}+R_{\alpha
}(n)\Bigr)n_{\alpha }-\sum_{\alpha \beta }\widetilde{n}_{\alpha }D_{\alpha
\beta }(n)\widetilde{n}_{\beta }\biggr)  \nonumber \\ &=&\int
dtd^{d}x\sum_{\alpha }\lambda _{\alpha }\widetilde{n}_{\alpha }
\biggl(\lambda _{\alpha }^{\,-1}\partial _{t}+\tau _{\alpha }-\nabla ^{2}+
\frac{1}{2}\Bigl(\sum_{\beta }g_{\alpha \beta }n_{\beta }-g_{\alpha }
\widetilde{n}_{\alpha }\Bigr)\biggr)n_{\alpha }\ .  \label{4}
\end{eqnarray}
In the second line we have neglected subleading terms. Correlation and
response functions can now be expressed as functional averages of
monomials of the $n_{\alpha }$ and $\widetilde{n}_{\alpha }$ with weight
$\exp (-{\cal J)}$. A glance at Eq.\ (\ref{PathInt}) shows that the
responses are defined with respect to additional local particle sources
$\widetilde{h}_{\alpha }\left( {\bf x},t\right) \geq 0$ in the Langevin
equations (\ref{1}). A rescaling of the fields $n_{\alpha }\rightarrow
l_{\alpha }n_{\alpha }$, $ \widetilde{n}_{\alpha }\rightarrow l_{\alpha
}^{\,-1}\widetilde{n}_{\alpha }$ leaves the functional ${\cal J}$
forminvariant but transforms the coupling constants $g_{\alpha \beta
}\rightarrow l_{\beta }^{\,}g_{\alpha \beta }$ and $g_{\alpha }\rightarrow
l_{\alpha }^{-1}g_{\alpha }$. Thus invariant coupling constants are given
by $f_{\alpha \beta }=g_{\alpha \beta }g_{\beta }$. A suitable rescaling
is defined by $g_{\alpha \alpha }=g_{\alpha }$. If we choose this
normalization, we denote the rescaled fields by $s_{\alpha }\sim n_{\alpha
}$ and $\widetilde{s}_{\alpha }\sim \widetilde{n}_{\alpha }$.

The scaling by a suitable mesoscopic length and time scale, $\mu ^{-1}$
and $\left(\lambda \mu ^{2}\right)^{-1}$ (with $\lambda_{\alpha}\sim
\lambda$) respectively, leads to $\widetilde{s}_{\alpha }\sim s_{\alpha
}\sim \mu ^{d/2}$, $f_{\alpha \beta }\sim \mu ^{\varepsilon }$ where
$\varepsilon =4-d$ . Hence $d_{c}=4$ is the upper critical dimension. It
is now easy to show that all neglected possible subleading terms in the
expansion of $R_{\alpha }(n)$ and the noise correlation, as well as higher
gradient terms in the Langevin equation\ (\ref{1}) and non-Gaussian and
non-Markovian noise correlations, have coupling constants with negative
$\mu $-dimensions near the upper critical dimension. Therefore, under the
renormalization group flow, they are renormalized to zero and we can
safely neglect them because we are interested in the leading universal
critical behavior. These couplings can be reintroduced if one is
interested in corrections to scaling. In a renormalized field theory
context \cite{Am84,ZiJu93}, the $\mu $-dimensions of the coupling
constants are equal to the so called naive or engineering dimensions. All
coupling constants with positive naive dimensions (the relevant couplings
with respect to the Gaussian fixed point as the starting point of the
perturbation expansion) need a renormalization because the corresponding
vertex functions are the only ones which develop primitive divergencies in
perturbation theory. Thus the field theory based on the functional ${\cal
J}$ (Eq.\ (\ref{4})) is renormalizable and the calculated scaling
properties are universal for the full class of MDP-processes.

\section{Renormalization and one-loop calculation}

Now we are in a position to develop the perturbation theory in the
inactive phase with all the $\tau _{\alpha }>0$. To begin with we separate
the anharmonic ``interaction'' terms from the dynamic functional ${\cal
J}$, Eq.\ (\ref{4}), and retain only the harmonic ones:
\begin{equation}
{\cal J}_{0}=\int dtd^{d}x\sum_{\alpha }\biggl(\widetilde{s}_{\alpha
}\Bigl(\partial _{t}+\lambda _{\alpha }(\tau _{\alpha }-\nabla
^{2})\Bigr)s_{\alpha }-\widetilde{h}_{\alpha }\widetilde{s}_{\alpha
}-h_{\alpha }s_{\alpha }\biggr)\ .  \label{5}
\end{equation}
Here we have included the external sources $\widetilde{h}_{\alpha }$ and $
h_{\alpha }$. The Gaussian path integral $\int {\cal
D}[\widetilde{s},s]\exp (-{\cal J}_{0})=\exp ((h,G\widetilde{h}))$
involves only non-negative fields $s_{\alpha }({\bf x},t)\geq 0$ as long
as $\widetilde{h}_{\alpha }({\bf x} ,t)\geq 0$. It yields the propagators
for the Fourier transformed fields $s( {\bf q},t)=\int d^{d}x\,s({\bf
x},t)\exp (-i{\bf q}\cdot {\bf x})$ etc., as
\begin{eqnarray}
\langle s_{\alpha }({\bf q},t)\widetilde{s}_{\beta }({\bf q}^{\prime
},t^{\prime })\rangle _{0} &=&(2\pi )^{d}\delta ({\bf q}+{\bf q}^{\prime
})\delta _{\alpha ,\beta }G_{\alpha }({\bf q},t-t^{\prime })  \nonumber \\
G_{\alpha }({\bf q},t) &=&\theta (t)\exp \Bigl(-\lambda _{\alpha }(\tau
_{\alpha }+q^{2})t\Bigr)  \label{6}
\end{eqnarray}
where the Heaviside theta-function is defined with $\theta (t=0)=0$
following from the Ito-discretization of the path-integral and ensuring
causality.

Besides the propagators, the anharmonic coupling terms in ${\cal J}$, Eq.\
(\ref{4}), define the elements of the graphical perturbation expansion.
They are depicted in FIG.~1 where an arrow marks a $\widetilde{s}$-leg and
we draw diagrams with the arrows always directed to the left (ascending
time ordering from right to left).

\bigskip

\begin{center}
\epsfig{file=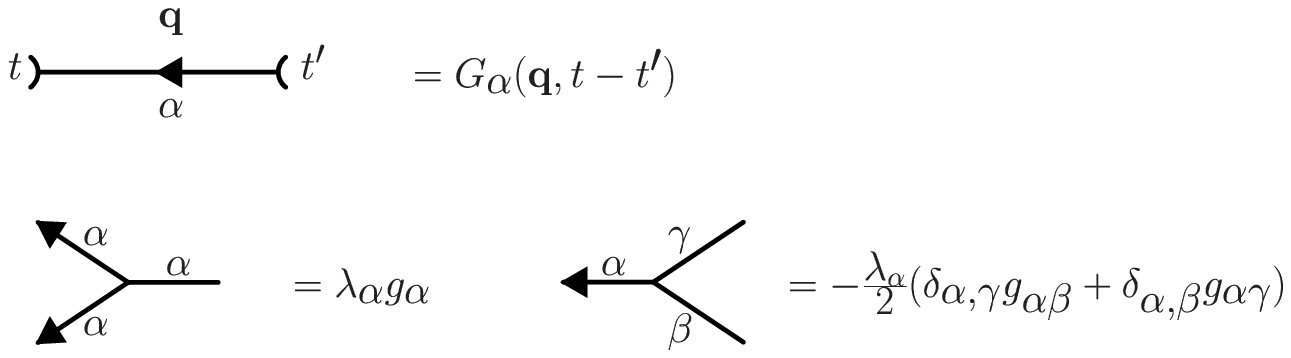,width=12cm}

FIG.\ 1. Elements of the graphical perturbation expansion
\end{center}

FIG.~1 shows that the color of a $\widetilde{s}$-leg on the left side of a
vertex is not annihilated: at least one $s$-leg on the right side displays
the same color. Thus, going from left to right, i.e.\ backward in time,
through a diagram, colors have only sources and no sinks. This property is
a consequence of the existence of absorbing states in the model. The
perturbation expansion of a translationally invariant field theory can be
analyzed by the calculation of the vertex functions $\Gamma _{\left\{
\alpha \right\} ,\left\{ \beta \right\} }(\left\{ {\bf q},\omega \right\}
)$ corresponding to the one-particle irreducible amputated diagrams. Here
the sets $\left\{ \alpha \right\} $ and $\left\{ \beta \right\} $ denote
the colors of the amputated outer $\widetilde{s}$-legs and $s$-legs
respectively. Going backward in time we conclude from the conservation
property of the $\widetilde{s}$-colors that the colors of the set $\left\{
\alpha \right\} $ must appear as a subset of $\left\{ \beta \right\} $.
Therefore the only nonzero two and three point vertex functions are
$\Gamma _{\alpha ,\alpha }$, $\Gamma _{\alpha \alpha ,\alpha }$, and
$\Gamma _{\alpha ,\alpha \beta }$. Moreover, another property follows
directly from color conservation: the vertex function $\Gamma _{\left\{
\alpha \right\} ,\left\{ \beta \right\} }$ does not depend on parameters
$\lambda $ , $\tau $, and $g$ with colors other than the ones of the
vertex functions itself. Thus, $\Gamma _{\alpha ,\alpha }$, $\Gamma
_{\alpha \alpha ,\alpha }$ , and $\Gamma _{\alpha ,\alpha \alpha }$ are
only functions of the parameters $\lambda _{\alpha }$, $\tau _{\alpha }$,
and $g_{\alpha \alpha }=g_{\alpha }$ and are in particular independent
from the interspecies couplings $g_{\alpha \beta }$ with $\alpha \neq
\beta $. Therefore these vertex functions are the same as the
corresponding functions of the well analyzed one-species Gribov process.
To calculate them one can set the interspecies couplings $g_{\alpha \beta
}=0 $. In this case the model shows rapidity reversal invariance
$\widetilde{s}_{\alpha }(t)\leftrightarrow -s_{\alpha }(-t)$ from which
follows the equality $\Gamma _{\alpha \alpha ,\alpha }=-\Gamma _{\alpha
,\alpha \alpha }$.

Now we are ready to consider the renormalization of the model. It is known
that the perturbation expansion of a field theory with a momentum cutoff $
\Lambda $ develops divergencies if $\Lambda \rightarrow \infty $ \cite
{Am84,ZiJu93}. If the model is renormalizable one can absorb all these
``primitive divergencies'' order by order in a loop expansion in a
suitable reparametrization of the parameters. Absorbing the primitive
divergencies regularizes the full model. The primitively divergent vertex
functions have nonnegative naive $\mu $ -dimensions. Here, they are
$\Gamma _{\alpha ,\alpha }\sim \mu ^{2}$ and $\Gamma _{\alpha \alpha
,\alpha },$ $\Gamma _{\alpha ,\alpha \beta }\sim \mu ^{\varepsilon }$,
logarithmic at the upper critical dimension, (FIG.~2).

\bigskip

\begin{center}
\epsfig{file=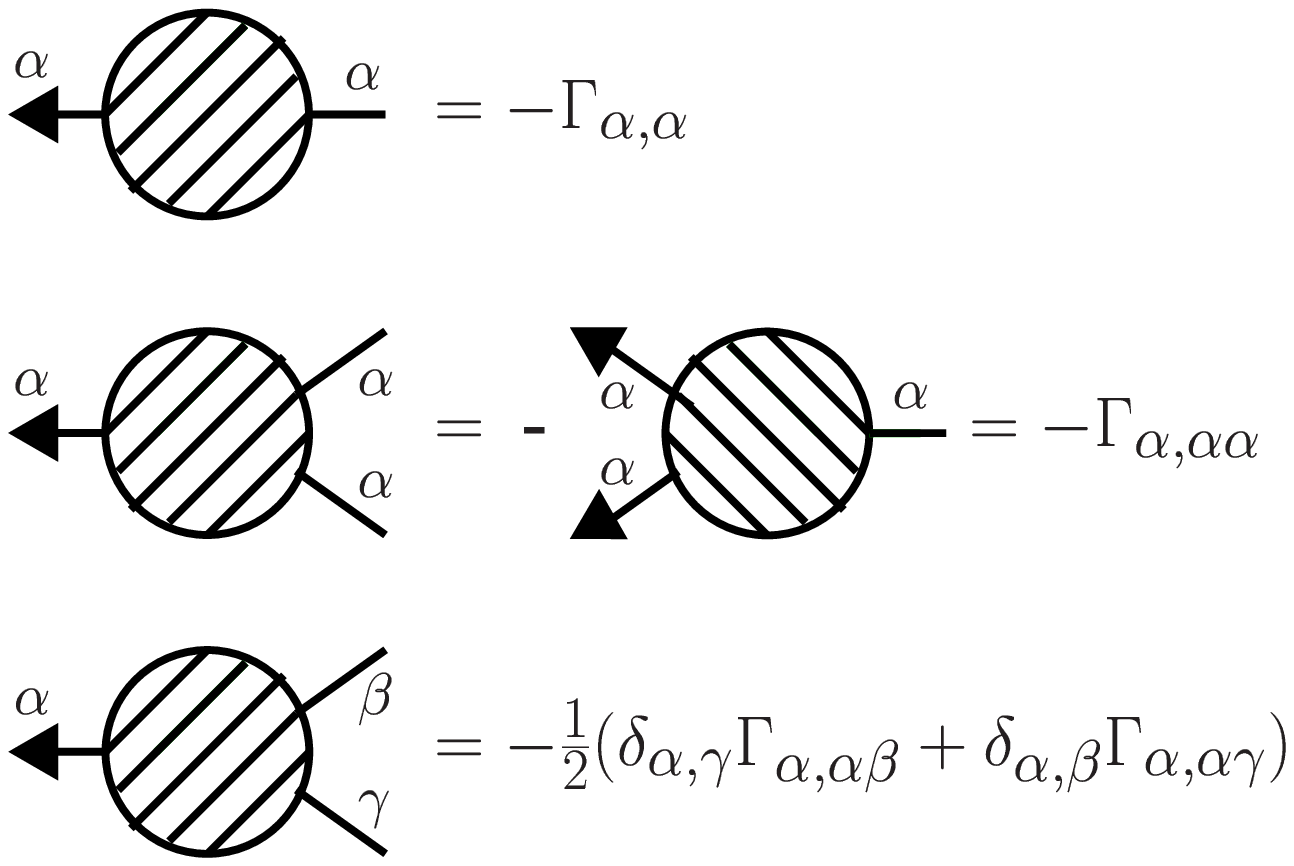,width=8cm}

FIG.\ 2. Primitively divergent vertex functions
\end{center}

Taking into account the general properties of the vertex functions found in
the foregoing paragraph, we see that the following renormalization scheme
renders the theory finite
\begin{eqnarray}
s_{\alpha } &\rightarrow &\mathaccent"7017{s}_{\alpha
}=\sqrt{Z_{s}^{(\alpha )}}\,s_{\alpha }\ ,\quad \widetilde{s}_{\alpha
}\rightarrow \mathaccent"7017{ \ \tilde{s}}_{\alpha }=\sqrt{Z_{s}^{(\alpha
)}}\,\widetilde{s}_{\alpha }\ ,\quad \lambda _{\alpha }\rightarrow
\mathaccent"7017{\lambda }_{\alpha }= \frac{Z_{\lambda }^{(\alpha
)}}{Z_{s}^{(\alpha )}}\,\lambda _{\alpha }\ , \nonumber \\[0.25cm] \tau
_{\alpha } &\rightarrow &\mathaccent"7017{\tau }_{\alpha }=\frac{ Z_{\tau
}^{(\alpha )}}{Z_{\lambda }^{(\alpha )}}{\tau }_{\alpha }\ ,\quad
f_{\alpha \beta }\rightarrow \mathaccent"7017{f}_{\alpha \beta
}=G_{\varepsilon }^{\,-1}\mu ^{\varepsilon }\,\frac{Z_{u}^{(\alpha \beta
)}}{ Z_{s}^{(\beta )}Z_{\lambda }^{(\alpha )}Z_{\lambda }^{(\beta
)}}\,u_{\alpha \beta }\ .  \label{7}
\end{eqnarray}
Here $G_{\varepsilon }=\Gamma \left( 1+\varepsilon /2\right) /\left( 4\pi
\right) ^{d/2}$ is a convenient constant. Instead of a momentum-cutoff
regularization, we use dimensional regularization and minimal
renormalization in the following. From the discussion above we learn that
the renormalization factors $Z_{i}^{(\alpha )}$ with $i=\left( s,\tau
,\lambda \right) $ and $Z_{u}^{(\alpha )}:=Z_{u}^{(\alpha \alpha )}$,
which are determined from $\Gamma _{\alpha ,\alpha }$ and $\Gamma _{\alpha
\alpha ,\alpha }$, are already known from the one-species Gribov process
and depend only on $u_{\alpha }:=u_{\alpha \alpha }$. Thus $Z_{i}^{(\alpha
)}=Z_{i}(u_{\alpha },\varepsilon)$. The new renormalization factors $
Z_{u}^{(\alpha \beta )}=Z_{int}(\left\{ u\right\} ,\left\{ \lambda
\right\} ,\varepsilon )$ with $\alpha \neq \beta $ stem from the
interspecies couplings. They depend only on the couplings $u_{\alpha }$,
$u_{\beta }$, $ u_{\alpha \beta }$, $u_{\beta \alpha }$, and the ratio
$\lambda _{\alpha }/\lambda _{\beta }$.

We will now explicitly calculate the renormalizations to one-loop order.
The primitively divergent one-loop diagrams are shown in (FIG.~3).

\bigskip

\begin{center}
\epsfig{file=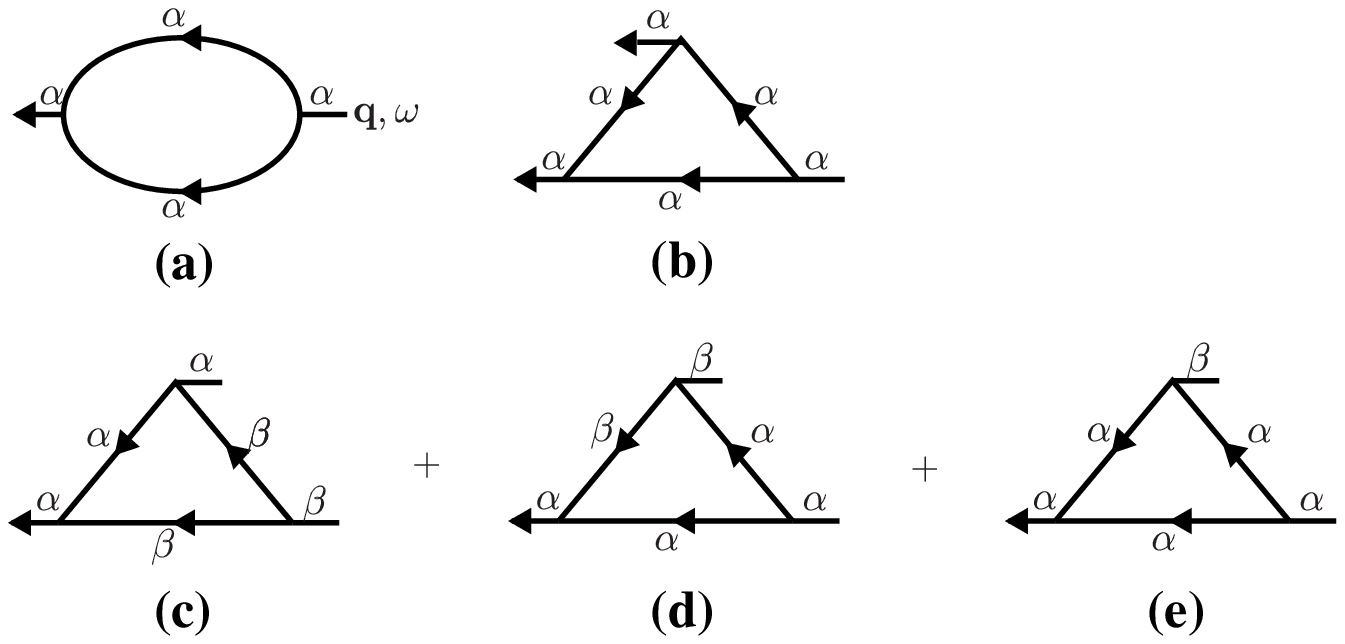,width=12cm}

FIG.\ 3. Primitively divergent one-loop diagrams
\end{center}

Using dimensional regularization, we express the contribution of the
self-energy diagram, FIG.\ 3(a), as a function of external momentum and
frequency, ${\bf q}$ and $\omega $:
\begin{eqnarray}
3(a) &=&-\frac{(\lambda _{\alpha }g_{\alpha })^{2}}{2}\int_{{\bf
p}}\frac{1}{ i\omega +2\lambda _{\alpha }\tau _{\alpha }+\lambda _{\alpha
}{\bf p} ^{2}+\lambda _{\alpha }({\bf p+q)}^{2}}  \nonumber \\
&=&\frac{G_{\varepsilon }}{4\varepsilon }\tau _{\alpha }^{-\varepsilon
/2}\lambda _{\alpha }g_{\alpha }^{\,2}\biggl(\frac{4\tau _{\alpha }}{
2-\varepsilon }+\frac{i\omega }{\lambda _{\alpha }}+\frac{{\bf q}^{2}}{2}
\biggr)\ .  \label{8}
\end{eqnarray}
Here, we have retained only terms linear in $\omega $ and ${\bf q}^{2}$.
These are the terms that display poles in $\varepsilon =4-d>0$.

To determine the primitive divergencies of the vertex functions we can set
external momenta and frequencies to zero and use equal temperatures $\tau
_{\alpha }=\tau >0$ as infrared regularisators. The contributions of the
three diagrams FIG.~3(c,d,e) add to
\begin{eqnarray}
3(c)+3(d)+3(e) &=&\frac{\lambda _{\alpha }g_{\alpha \beta
}}{2}\biggl(\frac{ \lambda _{\alpha }g_{\alpha \beta }}{2}\lambda _{\beta
}g_{\beta }\int_{{\bf p}}\frac{1}{2\lambda _{\beta }(\lambda _{\alpha
}+\lambda _{\beta })(\tau + {\bf p}^{2})^{2}}  \nonumber \\
&&+\frac{\lambda _{\beta }g_{\beta \alpha }}{2}\lambda _{\alpha }g_{\alpha
}\int_{{\bf p}}\frac{1}{2\lambda _{\alpha }(\lambda _{\alpha }+\lambda
_{\beta })(\tau +{\bf p}^{2})^{2}}  \nonumber \\ &&+\lambda _{\alpha
}^{\,2}g_{\alpha \alpha }g_{\alpha }\int_{{\bf p}}\frac{1 }{(2\lambda
_{\alpha })^{2}(\tau +{\bf p}^{2})^{2}}\biggr)  \nonumber \\
&=&\frac{G_{\varepsilon }}{4\varepsilon }\tau ^{-\varepsilon /2}\lambda
_{\alpha }g_{\alpha \beta }\biggl(\frac{\lambda _{\alpha }g_{\alpha \beta
}g_{\beta }+\lambda _{\beta }g_{\beta \alpha }g_{\alpha }}{\lambda
_{\alpha }+\lambda _{\beta }}+g_{\alpha \alpha }g_{\alpha }\biggr)\ .
\label{9}
\end{eqnarray}
From the zero-loop contributions and the results of our short calculation,
Eqs.\ (\ref{8},\ref{9}), we obtain the (as yet unrenormalized) one-loop
vertex functions to the desired order in $\omega $ and ${\bf q}^{2}$:
\begin{eqnarray}
\Gamma _{\alpha \alpha } &=&i\omega \biggl(1-\frac{G_{\varepsilon }}{
4\varepsilon }g_{\alpha \alpha }g_{\alpha }\tau _{\alpha }^{-\varepsilon
/2} \biggr)+\lambda _{\alpha }{\bf q}^{2}\biggl(1-\frac{G_{\varepsilon }}{
8\varepsilon }g_{\alpha \alpha }g_{\alpha }\tau _{\alpha }^{-\varepsilon
/2} \biggr)  \nonumber \\ &&+\lambda _{\alpha }\tau _{\alpha
}\biggl(1-\frac{G_{\varepsilon }}{ 2\varepsilon (1-\varepsilon
/2)}g_{\alpha \alpha }g_{\alpha }\tau _{\alpha }^{-\varepsilon /2}\biggr)
\label{10}
\end{eqnarray}
and
\begin{equation}
\Gamma _{\alpha ,\alpha \beta }=\lambda _{\alpha }g_{\alpha \beta
}\Biggl(1- \frac{G_{\varepsilon }}{2\varepsilon }\biggl(\frac{\lambda
_{\alpha }g_{\alpha \beta }g_{\beta }+\lambda _{\beta }g_{\beta \alpha
}g_{\alpha }}{ \lambda _{\alpha }+\lambda _{\beta }}+g_{\alpha \alpha
}g_{\alpha }\biggr) \tau ^{-\varepsilon /2}\Biggr)\ .  \label{11}
\end{equation}
An explicit calculation of diagram (b) of FIG.~3 demonstrates that indeed
$ \Gamma _{\alpha \alpha ,\alpha }=-\Gamma _{\alpha ,\alpha \alpha }$ if $
g_{\alpha \alpha }=g_{\alpha }$.

To absorb the $\varepsilon $-poles in the renormalization $Z$-factors we
note that the vertex functions are renormalized by the scheme Eq.\ (\ref{7})
as
\begin{equation}
\Gamma _{\alpha _{1}\cdots \alpha _{n}}\rightarrow \mathaccent"7017\Gamma
_{\alpha _{1}\cdots \alpha _{n}}=\Bigl(Z_{s}^{(\alpha _{1})}\cdots
Z_{s}^{(\alpha _{n})}\Bigr)^{-1/2}\Gamma _{\alpha _{1}\cdots \alpha _{n}}\ .
\label{12}
\end{equation}
Using again the renormalization scheme Eq.\ (\ref{7}), we find the
renormalized vertex functions from Eqs.\ (\ref{10},\ref{11}) as
\begin{eqnarray}
\Gamma _{\alpha \alpha } &=&i\omega \biggl(Z_{s}^{(\alpha
)}-\frac{u_{\alpha }}{4\varepsilon }(\mu /\tau _{\alpha })^{\varepsilon
/2}\biggr)+\lambda _{\alpha }{\bf q}^{2}\biggl(Z_{\lambda }^{(\alpha
)}-\frac{u_{\alpha }}{ 8\varepsilon }(\mu /\tau _{\alpha })^{\varepsilon
/2}\biggr)  \nonumber \\ &&+\lambda _{\alpha }\tau _{\alpha
}\biggl(Z_{\tau }^{(\alpha )}-\frac{ u_{\alpha }}{2\varepsilon
(1-\varepsilon /2)}(\mu /\tau _{\alpha })^{\varepsilon /2}\biggr)
\label{13}
\end{eqnarray}
and
\begin{equation}
\Gamma _{\alpha ,\alpha \beta }\Gamma _{\beta ,\beta \beta
}=G_{\varepsilon }^{-1}\mu ^{\varepsilon }\lambda _{\alpha
}^{\,2}u_{\alpha \beta }\Biggl( Z_{u}^{(\alpha \beta
)}-\frac{1}{2\varepsilon }\biggl(\frac{\lambda _{\alpha }u_{\alpha \beta
}+\lambda _{\beta }u_{\beta \alpha }}{\lambda _{\alpha }+\lambda _{\beta
}}+u_{\alpha }+2u_{\beta }\biggr)(\mu /\tau )^{\varepsilon /2}\Biggr)\ .
\label{14}
\end{equation}
Therefore the vertex functions become finite by choosing
\begin{eqnarray}
Z_{s}^{(\alpha )} &=&1+\frac{u_{\alpha }}{4\varepsilon }\ ,\qquad Z_{\lambda
}^{(\alpha )}=1+\frac{u_{\alpha }}{8\varepsilon }\ ,\qquad Z_{\tau
}^{(\alpha )}=1+\frac{u_{\alpha }}{2\varepsilon }\ ,  \nonumber \\
Z_{u}^{(\alpha \beta )} &=&1+\frac{1}{2\varepsilon }\biggl(\frac{\lambda
_{\alpha }u_{\alpha \beta }+\lambda _{\beta }u_{\beta \alpha }}{\lambda
_{\alpha }+\lambda _{\beta }}+u_{\alpha }+2u_{\beta }\biggr)  \label{15}
\end{eqnarray}
up to higher orders in the coupling constants $u$. As anticipated, for $
\alpha =\beta $ we have found the well known renormalization factors of
the Reggeon field theory.

\section{Renormalization Group Analysis and Asymptotic Scaling}

Next we will explore the scaling properties of multicolored directed
percolation. Scaling properties describe how physical quantities will
transform under a change of length scales. At the end of chapter II we
have introduced the arbitrary mesoscopic length scale $\mu $. The freedom
in the choice of $\mu $, keeping the unrenormalized bare parameters
$\left\{ \mathaccent"7017{\tau }_{\alpha },\mathaccent"7017{\lambda
}_{\alpha }, \mathaccent"7017{g}_{\alpha },\mathaccent"7017{g}_{\alpha
\beta }\right\} $, and -- in cutoff regularization -- the momentum cutoff
$\Lambda $ fixed, can be used to derive in a routine fashion the
renormalization group (RG) equation for the connected correlation and
response functions, the Green functions
\begin{equation}
G^{\{N,\widetilde{N}\}}(\{{\bf x},t\})=\biggl\langle\prod_{\alpha }\biggl(
\prod_{i=1}^{N_{\alpha }}s({\bf x}_{i}^{(\alpha )},t_{i}^{(\alpha
)})\prod_{j=N_{\alpha }+1}^{N_{\alpha }+\widetilde{N}_{\alpha
}}\widetilde{s} ({\bf x}_{j}^{(\alpha )},t_{j}^{(\alpha
)})\biggr)\biggr\rangle^{conn}\ . \label{16}
\end{equation}
We denote $\mu $-derivatives at fixed bare parameters by $\left. \partial
_{\mu }\right| _{0}$. From $\mu \left. \partial _{\mu }\right| _{0}
\mathaccent"7017G^{\{N,\widetilde{N}\}}=0$ and the renormalization scheme
Eq.\ (\ref{7}), which leads to $\mathaccent"7017G^{\{N,\widetilde{N}
\}}=\prod_{\alpha }(Z_{s}^{(\alpha )})^{(N_{\alpha }+\widetilde{N}_{\alpha
})/2}\,G^{\{N,\widetilde{N}\}}$, we then find the RG equations
\begin{equation}
\biggl[{\cal D}_{\mu }+\sum_{\alpha }\frac{N_{\alpha }+\widetilde{N}_{\alpha
}}{2}\gamma _{s}^{(\alpha )}\biggr]G^{\{N,\widetilde{N}\}}=0  \label{17}
\end{equation}
with the renormalization group differential operator
\begin{equation}
{\cal D}_{\mu }=\mu \partial _{\mu }+\sum_{\alpha }\Bigl(\zeta _{\alpha
}\lambda _{\alpha }\partial _{\lambda _{\alpha }}+\kappa _{\alpha }\tau
_{\alpha }\partial _{\tau _{\alpha }}+\beta _{\alpha }\partial _{u_{\alpha
}}\Bigr)+\sum_{\alpha \neq \beta }\beta _{\alpha \beta }\partial _{u_{\alpha
\beta }}\ .  \label{18}
\end{equation}
Here we have introduced the Gell-Mann--Low functions
\begin{eqnarray}
\zeta _{\alpha } &=&\mu \left. \partial _{\mu }\right| _{0}\ln \lambda
_{\alpha }=\gamma _{s}^{(\alpha )}-\gamma _{\lambda }^{(\alpha )}\ ,
\nonumber \\
\kappa _{\alpha } &=&\mu \left. \partial _{\mu }\right| _{0}\ln \tau
_{\alpha }=\gamma _{\lambda }^{(\alpha )}-\gamma _{\tau }^{(\alpha )}\ ,
\nonumber \\
\beta _{\alpha \beta } &=&\mu \left. \partial _{\mu }\right| _{0}u_{\alpha
\beta }=(-\varepsilon +\gamma _{s}^{(\beta )}+\gamma _{\lambda }^{(\alpha
)}+\gamma _{\lambda }^{(\beta )}-\gamma _{u}^{(\alpha \beta )})u_{\alpha
\beta }\   \label{19}
\end{eqnarray}
with $\beta _{\alpha }=\beta _{\alpha \alpha }$, and the Wilson functions
\begin{equation}
\gamma _{i}^{(r)}=\mu \left. \partial _{\mu }\right| _{0}\ln Z_{i}^{(r)}\
,\qquad i=s,\lambda ,\tau ,u,\qquad r=\alpha ,\alpha \beta \ .  \label{20}
\end{equation}
The RG equations\ (\ref{17}) can be solved in terms of a single flow
parameter $l$ using characteristics. Following this method, flowing
parameters are defined by the characteristic equations
\begin{eqnarray}
l\frac{d}{dl}\bar{u}_{\alpha \beta }(l) &=&\beta _{\alpha \beta }(\bar{u}
(l))\ ,\qquad \qquad \qquad \qquad \qquad \bar{u}_{\alpha \beta
}(1)=u_{\alpha \beta }\ ,  \nonumber \\ l\frac{d}{dl}\bar{\lambda}_{\alpha
}(l) &=&\zeta _{\alpha }(\bar{u}(l))\bar{ \lambda}_{\alpha }(l)=\zeta
(\bar{u}_{\alpha }(l))\bar{\lambda}_{\alpha }(l)\ ,\qquad
\bar{\lambda}_{\alpha }(1)=\lambda _{\alpha }\ ,  \nonumber \\
l\frac{d}{dl}\bar{\tau}_{\alpha }(l) &=&\kappa _{\alpha }(\bar{u}(l))\bar{
\tau}_{\alpha }(l)=\kappa (\bar{u}_{\alpha }(l))\bar{\tau}_{\alpha }(l)\
,\qquad \bar{\tau}_{\alpha }(1)=\tau _{\alpha }\ ,  \label{21a}
\end{eqnarray}
and the RG equations\ (\ref{17}) of the Green functions become
\begin{equation}
\biggl[l\frac{d}{dl}+\sum_{\alpha }\frac{N_{\alpha }+\widetilde{N}_{\alpha
} }{2}\gamma (\bar{u}_{\alpha }(l))\biggr]G^{\{N,\widetilde{N}\}}(\{{\bf
x} ,t\},\{\bar{\tau}(l)\},\{\bar{u}(l)\},\{\bar{\lambda}(l)\},l\mu )=0\ .
\label{21b}
\end{equation}
Here, the functions $\gamma =\gamma _{s}^{(\alpha )}$, $\zeta =\zeta
_{\alpha }$, and $\kappa =\kappa _{\alpha }$ are independent of the color
$ \alpha $ and the interspecies couplings. The flow equations\ (\ref{21a}
) describe how the parameters transform if we change the momentum scale
$\mu $ according to $\mu \rightarrow \bar{\mu}(l)=l\mu $. Being interested
in the infrared (IR) behavior of the theory, we must study the limit
$l\rightarrow 0 $. In general we expect that in this IR limit the coupling
constants $\bar{ u}_{\alpha \beta }(l)$ flow to a stable fixed point
$u_{\alpha \beta ,*}$ according to the first set of Eq.\ (\ref{21a}). In
particular, the intraspecies couplings $\bar{u}_{\alpha \alpha
}=\bar{u}_{\alpha }$ then flow to a color independent fixed point
$\bar{u}_{\alpha }(0)=u_{*}$ because $\beta _{\alpha }(\bar{u})=\beta
(\bar{u}_{\alpha })$ , and this Gell-Mann--Low function $\beta $ is equal
to the corresponding function known from the one-species Gribov process.
Thus, the fixed point value $ u_{*} $ is independent from any coupling to
other species.

The solutions of the second and third set of the flow equations\ (\ref{21b})
are readily found in terms of the functions $\bar{u}_{\alpha }(l)$. In the
IR limit $l\ll 1$ they have the scaling form
\begin{equation}
\bar{\lambda}_{\alpha }(l)=l^{z-2}A_{\lambda }^{(\alpha )}\lambda _{\alpha
}\ ,\qquad \bar{\tau}_{\alpha }(l)=l^{2-1/\nu }A_{\tau }^{(\alpha )}\tau
_{\alpha }  \label{22}
\end{equation}
where the $A_{i}^{(\alpha )}$ are nonuniversal amplitude factors. The
scaling exponents
\begin{equation}
z=2+\zeta (u_{*})\ ,\qquad \nu =\frac{1}{2-\kappa (u_{*})}  \label{23}
\end{equation}
are already known from directed percolation. From Eq.\ (\ref{21b}) we find
the solution in the IR limit as
\begin{eqnarray}
G^{\{N,\widetilde{N}\}}(\{{\bf x},t\},\{\tau \},\{u\},\{\lambda \},\mu )
&=&l^{(N+\widetilde{N})\eta /2}\prod_{\alpha }(A_{s}^{(\alpha
)})^{(N_{\alpha }+\widetilde{N}_{\alpha })/2}  \nonumber \\
&&\hspace{-2cm}\times G^{\{N,\widetilde{N}\}}(\{{\bf x},t\},\{l^{2-1/\nu
}A_{\tau }\tau \},\{u_{*}\},\{l^{z-2}A_{\lambda }\lambda \},l\mu )
\label{24}
\end{eqnarray}
with $N=\sum_{\alpha }N_{\alpha }$, $\widetilde{N}=\sum_{\alpha
}\widetilde{N }_{\alpha }$, and the DP anomalous field scaling exponent
\begin{equation}
\eta =\gamma (u_{*})\ .  \label{25}
\end{equation}
Omitting the nonuniversal amplitude factors $A_{i}^{(\alpha )}$ and taking
into account dimensional analysis in space and time
\begin{equation}
G^{\{N,\widetilde{N}\}}(\{{\bf x},t\},\{\tau \},\{u\},\{\lambda \},\mu )=\mu
^{(N+\widetilde{N})d/2}G^{\{N,\widetilde{N}\}}(\{\mu {\bf x},\lambda \mu
^{2}t\},\{\tau /\mu ^{2}\},\{u\},\{c\},1)  \label{26}
\end{equation}
where $c_{\alpha }=\lambda _{\alpha }/\lambda $, we get by combination of
Eqs.\ (\ref{24},\ref{26}) the asymptotic scaling form of the Green functions
\begin{equation}
G^{\{N,\widetilde{N}\}}(\{{\bf x},t\},\{\tau \})=l^{(N+\widetilde{N})(d+\eta
)/2}G^{\{N,\widetilde{N}\}}(\{l{\bf x},l^{z}t\},\{\tau /l^{1/\nu }\})\ .
\label{27}
\end{equation}
This has the important consequence that all scaling properties of the DP
processes remain unaffected by the introduction of many colors. Moreover,
all intraspecies Green functions are completely independent of the
coupling between the species, which follows from the absorbing state
conditions for each color.

However, the interspecies coupling constants $u_{\alpha \beta }$ with $
\alpha \neq \beta $ determine the properties of the interspecies scaling
functions. Therefore we will now consider the consequences of the flow
equations for these parameters. For this purpose we need the
Gell-Mann--Low functions $\beta _{\alpha \beta }$ explicitly. From the
last of Eqs.\ (\ref {19}) we know that each of these functions begins with
the zero-loop term $ -\varepsilon u_{\alpha \beta }$, and the higher order
terms are determined by the Wilson functions. These functions, the
logarithmic derivatives of the $Z$ -factors, are given by $\gamma =\mu
\left. \partial _{\mu }\right| _{0}\ln Z=\sum_{\alpha \beta }\beta
_{\alpha \beta }\partial _{u_{\alpha \beta }}\ln Z$. In minimal
renormalization the $Z$-factors have a pure Laurent expansion with respect
to $\varepsilon $: $Z=1+Y^{(1)}/\varepsilon +Y^{(2)}/\varepsilon
^{2}+\cdots $. Thus recursively in the loop expansion the Wilson functions
also have a pure Laurent expansion and, because they are finite for
$\varepsilon \rightarrow 0$, this expansion reduces to the constant term,
i.e.\ all $\varepsilon $-poles have to cancel in the logarithmic
derivation. Hence, we obtain the Wilson functions simply from the formula
$\gamma =-\sum_{\alpha \beta }u_{\alpha \beta }\partial _{u_{\alpha \beta
}}Y^{(1)}$. Now it is easy to get these functions from the one loop
results of the $Z$-factors Eqs.\ (\ref{15}). We find to this order
\begin{eqnarray}
\gamma _{s}^{(\alpha )} &=&-\frac{u_{\alpha }}{4}\ ,\qquad \gamma
_{\lambda }^{(\alpha )}=-\frac{u_{\alpha }}{8}\ ,\qquad \gamma _{\tau
}^{(\alpha )}=- \frac{u_{\alpha }}{2}\ ,  \nonumber \\ \gamma
_{u}^{(\alpha \beta )} &=&-\frac{1}{2}\biggl(\frac{\lambda _{\alpha
}u_{\alpha \beta }+\lambda _{\beta }u_{\beta \alpha }}{\lambda _{\alpha
}+\lambda _{\beta }}+u_{\alpha }+2u_{\beta }\biggr)\ ,  \label{28}
\end{eqnarray}
from which one finds the Gell-Mann--Low functions Eq.\ (\ref{19}) as
\begin{equation}
\beta _{\alpha \beta }=\biggl(-\varepsilon +\frac{3u_{\alpha }}{8}+\frac{
5u_{\beta }}{8}+\frac{\lambda _{\alpha }u_{\alpha \beta }+\lambda _{\beta
}u_{\beta \alpha }}{2(\lambda _{\alpha }+\lambda _{\beta
})}\biggr)u_{\alpha \beta }\ .  \label{29}
\end{equation}

The Gell-Mann--Low functions of the intraspecies couplings, the
``diagonal'' $\beta $-functions, follow as $\beta _{\alpha }=(-\varepsilon
+3u_{\alpha }/2)u_{\alpha }$ giving the stable fixed point values
$u_{\alpha *}=u_{*}=2\varepsilon /3$. With the help of Eqs.\
(\ref{19},\ref{23},\ref {25}), this stable fixed point leads to the well
known one loop order DP exponents $z=2-\varepsilon /12$, $\nu
=1/2+\varepsilon /16$, and $\eta =-\varepsilon /6$ . Using the fixed point
values $u_{\alpha *}$ to obtain the Gell-Mann--Low functions of the
interspecies couplings of a pair of colors $\alpha \neq \beta $, we get
\begin{equation}
\beta _{\alpha \beta }=\biggl(-\frac{\varepsilon }{3}+\frac{\lambda _{\alpha
}u_{\alpha \beta }+\lambda _{\beta }u_{\beta \alpha }}{2(\lambda _{\alpha
}+\lambda _{\beta })}\biggr)u_{\alpha \beta }\ .  \label{30}
\end{equation}
In addition to the unstable decoupled fixed point values $u_{\alpha \beta
*}=u_{\beta \alpha *}=0$, the equations $\beta _{\alpha \beta }=\beta
_{\beta \alpha }=0$ are solved by a fixed point line
\begin{equation}
\frac{2\lambda _{\alpha }}{\lambda _{\alpha }+\lambda _{\beta }}u_{\alpha
\beta *}+\frac{2\lambda _{\beta }}{\lambda _{\alpha }+\lambda _{\beta }}
u_{\beta \alpha *}=\frac{4\varepsilon }{3}\ .  \label{31}
\end{equation}
This equation is the key result of this Section. Clearly, however, the
one-loop calculation cannot give us any information whether the degeneracy
of all the points on this line is fundamental to our model or is lifted by
higher loop corrections. Thus, we must proceed to two-loop order,
presented in the following chapter. We note, however, that there are two
special points of unidirectional coupling on the fixed line:
\begin{eqnarray}
u_{\alpha \beta *} &=&0\ ,\qquad u_{\beta \alpha *}=\frac{2(\lambda _{\alpha
}+\lambda _{\beta })}{3\lambda _{\beta }}\varepsilon \qquad \text{ and}
\nonumber \\
u_{\beta \alpha *} &=&0\ ,\qquad u_{\alpha \beta *}=\frac{2(\lambda _{\alpha
}+\lambda _{\beta })}{3\lambda _{\alpha }}\varepsilon \ .  \label{32}
\end{eqnarray}
In the next chapter we will show that for colors with the same flavor,
meaning $\lambda _{\alpha }=\lambda _{\beta }$, these two points are
indeed the only stable points on the line. Because $u_{\alpha \beta *}=0$
or $ u_{\beta \alpha *}=0$ are fixed point values at any loop order, we
conjecture that unidirectionality is generic for the asymptotic behavior
of coupled DP processes, irrespective of whether colors belong to
different flavors or not.

\section{Two-loop Results and Crossover to Unidirectionality}

From the analysis in the foregoing chapters we know that the interaction
of two colors does not depend on the existence of other ones. Thus in this
chapter we consider a coupled model of two colors $\alpha =1$ and $2$ with
the same flavor, i.e.\ equal kinetic coefficients $\lambda _{\alpha
}=\lambda $ and intraspecies couplings $g_{\alpha }=g$. In contrast, the
temperatures $\tau _{\alpha }$ may be different. Thus, the
(unrenormalized) dynamic functional is now given by
\begin{eqnarray}
{\cal J} &=&\int dtd^{d}x\,\lambda \biggl(\widetilde{s}_{1}\Bigl(\lambda
^{-1}\partial _{t}+\tau _{1}-\nabla ^{2}+\frac{g}{2}\bigl(s_{1}-\widetilde{s}
_{1}\bigr)\Bigr)s_{1}  \nonumber \\
&&+\widetilde{s}_{2}\Bigl(\lambda ^{-1}\partial _{t}+\tau _{2}-\nabla ^{2}+
\frac{g}{2}\bigl(s_{2}-\widetilde{s}_{2}\bigr)\Bigr)s_{2}+\frac{1}{2}\Bigl(
\widetilde{s}_{1}g_{12}+\widetilde{s}_{2}g_{21}\Bigr)s_{1}s_{2}\biggr)\;.
\label{33}
\end{eqnarray}
Note that in the case $g_{12}=0$, the dynamics of species $1$ completely
decouples from species $2$ and vice versa. It follows that $\Gamma _{\alpha
,\alpha \beta }=0$, if $g_{\alpha \beta }=0$ and $\alpha \neq \beta $.

The detailed calculation of the two-loop contributions is presented in
Appendices A to C. Adding Eq.\ (\ref{A9a}) to the zero- and one-loop parts
of the selfenergy Eq.\ (\ref{10}), we get after renormalization, using
Eq.\ (\ref{12}) and the scheme Eq.\ (\ref{7}),
\begin{eqnarray}
\Gamma _{\alpha \alpha } &=&\lambda \tau _{\alpha }\Biggl(Z_{\tau }-\frac{
(\mu ^{2}/\tau _{\alpha })^{\varepsilon /2}u}{2\varepsilon (1-\varepsilon
/2) }\biggl(1+u\Bigl(\frac{2}{\varepsilon }-\frac{3}{16}\Bigr)\biggr)
+C_{\tau } \frac{(\mu ^{2}/\tau _{\alpha })^{\varepsilon
}u^{2}}{\varepsilon ^{2}} \Biggr)  \nonumber \\ &&+\lambda {\bf
q}^{2}\Biggl(Z_{\lambda }-\frac{(\mu ^{2}/\tau _{\alpha })^{\varepsilon
/2}u}{8\varepsilon }\biggl(1+u\Bigl(\frac{13}{8\varepsilon } -
\frac{3}{16}\Bigr)\biggr)+C_{q^{2}}\frac{(\mu ^{2}/\tau _{\alpha
})^{\varepsilon }u^{2}}{\varepsilon ^{2}}\Biggr)  \nonumber \\ &&+i\omega
\Biggl(Z_{s}-\frac{(\mu ^{2}/\tau _{\alpha })^{\varepsilon /2}u}{
4\varepsilon }\biggl(1+u\Bigl(\frac{7}{4\varepsilon }-\frac{3}{16}\Bigr)
\biggr)+C_{\omega }\frac{(\mu ^{2}/\tau _{\alpha })^{\varepsilon }u^{2}}{
\varepsilon ^{2}}\Biggr)+O(\varepsilon ^{0}u^{2})\ ,  \label{35}
\end{eqnarray}
where the constants $C_{i}$ are given in Eq.\ (\ref{A9b}), and
$u=u_{\alpha } $. Note that in comparison with Eq.\ (\ref{13}) now the
one-loop terms acquire renormalizations to $O(u)$ in order to be
consistent in $O(u^{2})$ of the perturbation expansion. These renormalized
one-loop terms are needed to compensate non primitive singular terms
proportional to $\ln (\mu ^{2}/\tau ) $ arising now by the $\varepsilon
$-expansion of $\Gamma _{\alpha \alpha }$. Of course only primitive
UV-divergencies, which means here $\tau$-independent ones, have to be
regularized by renormalization to make the theory finite. Eliminating the
$\varepsilon$-poles from Eq.\ (\ref{35}) by the $Z$-factors, we find
\begin{eqnarray}
Z_{s} &=&1+\frac{u}{4\varepsilon }+\frac{u^{2}}{32\varepsilon
}\biggl(\frac{ 7 }{\varepsilon }-3+\frac{9}{2}\ln
\frac{4}{3}\biggr)+O(u^{3})\;,  \nonumber
\\
Z_{\lambda } &=&1+\frac{u}{8\varepsilon }+\frac{u^{2}}{128\varepsilon }
\biggl(\frac{13}{\varepsilon }-\frac{31}{4}+\frac{35}{2}\ln \frac{4}{3}
\biggr)+O(u^{3})\;,  \nonumber \\ Z_{\tau } &=&1+\frac{u}{2\varepsilon
}+\frac{u^{2}}{2\varepsilon }\biggl( \frac{1}{\varepsilon
}-\frac{5}{16}\biggr)+O(u^{3})\;.  \label{36}
\end{eqnarray}

In the same way as for the selfenergy we find the other vertex functions.
Renormalizing Eq.\ (\ref{11}) and adding the two-loop contribution Eq.\ (\ref
{A22}) we get
\begin{eqnarray}
\Gamma _{\alpha ,\alpha \beta } &=&G_{\varepsilon }^{\,-1/2}\mu
^{\varepsilon /2}\lambda u_{\alpha \beta }u^{-1/2}\Biggl(Z_{u}^{(\alpha
\beta )}Z_{u}^{\,-1/2}+\frac{(\mu ^{2}/\tau _{\alpha })^{\varepsilon /2}}{
2\varepsilon }\biggl(\Bigl(1+\frac{2u}{\varepsilon }-\frac{3u}{16}+\frac{
u_{\alpha \beta }+u_{\beta \alpha }}{4\varepsilon }\Bigr)u  \nonumber \\
&&+\Bigl(1+\frac{3u}{2\varepsilon }-\frac{3u}{16}+\frac{u_{\alpha \beta
}+u_{\beta \alpha }}{2\varepsilon }\Bigr)\frac{u_{\alpha \beta }+u_{\beta
\alpha }}{2}\biggr)  \nonumber \\ &&+\frac{(\mu ^{2}/\tau _{\alpha
})^{\varepsilon }}{16\varepsilon }\biggl( \Bigl(\frac{8}{\varepsilon
}+1\Bigr)u^{2}+\Bigl(\frac{4}{\varepsilon }+\frac{ 1}{2}\Bigr)u{}u_{\beta
\alpha }+\Bigl(\frac{2}{\varepsilon } +1\Bigr)u_{\alpha \beta }u_{\beta
\alpha }  \nonumber \\ &&+\Bigl(\frac{4}{\varepsilon }+\frac{3}{2}-3\ln
\frac{4}{3} \Bigr)u{}u_{\alpha \beta }+\Bigl(\frac{1}{\varepsilon
}+\frac{3}{2}\ln \frac{ 4}{3}\Bigr)\Bigl(u_{\alpha \beta }^{\,2}+u_{\beta
\alpha }^{\,2}\Bigr) \biggr) \Biggr)+O(\varepsilon ^{0}u^{2})\
\label{37}
\end{eqnarray}
where $Z_{u}=Z_{u}^{(\alpha \alpha )}$. Here the $\varepsilon $-expansion
also shows that the non primitive divergencies $\sim \ln (\mu ^{2}/\tau )$
cancel. We eventually find
\begin{eqnarray}
Z_{u}^{\left( \alpha \beta \right) } &=&1+\frac{1}{4\varepsilon }
\Bigl(6u+u_{\alpha \beta }+u_{\beta \alpha }\Bigr)  \nonumber \\[0.1cm]
&&+\frac{1}{16\varepsilon }\Biggl(\biggl(\frac{36}{\varepsilon
}-\frac{19}{2} \biggr)u^{2}+\biggl(\frac{8}{\varepsilon
}-\frac{5}{4}\biggr)uu_{\beta \alpha }+\biggl(\frac{2}{\varepsilon
}-1\biggr)u_{\alpha \beta }u_{\beta \alpha }  \nonumber \\
&&+\biggl(\frac{8}{\varepsilon }-\frac{9}{4}+3\ln \frac{4}{3}\biggr)
uu_{\alpha \beta }+\biggl(\frac{1}{\varepsilon }-\frac{3}{2}\ln
\frac{4}{3} \biggr)\Bigl(u_{\alpha \beta }^{2}+u_{\beta \alpha
}^{2}\Bigr)\Biggr) +O(u^{3})\ ,  \label{38}
\end{eqnarray}
and in particular
\begin{equation}
Z_{u}=1+\frac{2u}{\varepsilon }+\biggl(\frac{1}{\varepsilon }-\frac{1}{4}
\biggr)\frac{7u^{2}}{2\varepsilon }+O(u^{3})\ .  \label{39}
\end{equation}

We are now in a position to calculate the renormalization group functions
Eq.\ (\ref{19}) from Eqs.\ (\ref{36},\ref{38},\ref{39}) with the help of
Eq.\ (\ref{20}). They are given by
\begin{eqnarray}
\gamma &=&-\frac{u}{4}+\biggl(\frac{2}{3}-\ln \frac{4}{3}\biggr)\frac{9u^{2}
}{32}+O(u^{3})\ ,  \nonumber \\
\zeta &=&-\frac{u}{8}+\biggl(\frac{17}{2}-\ln \frac{4}{3}\biggr)\frac{u^{2}}{
128}+O(u^{3})\ ,  \nonumber \\
\kappa &=&\frac{3u}{8}-\biggl(\frac{7}{10}+\ln \frac{4}{3}\biggr)\frac{
35u^{2}}{128}+O(u^{3})\ ,  \label{40}
\end{eqnarray}
and
\begin{eqnarray}
\beta _{\alpha \beta } &=&\Biggl(-\varepsilon +u+\frac{1}{4}\Bigl(u_{\alpha
\beta }+u_{\beta \alpha }\Bigr)-\frac{1}{8}u_{\alpha \beta }u_{\beta \alpha
}-\frac{3}{16}\ln \frac{4}{3}\Bigl(u_{\alpha \beta }^{2}+u_{\beta \alpha
}^{2}\Bigr)  \nonumber \\
&&-\biggl(\frac{97}{106}+\ln \frac{4}{3}\biggr)\frac{53u^{2}}{64}-\biggl(
\frac{3}{4}-\ln \frac{4}{3}\biggr)\frac{3uu_{\alpha \beta }}{8}-\frac{5}{32}
uu_{\beta \alpha }+O(u^{3})\Biggr)u_{\alpha \beta }\ .  \label{40a}
\end{eqnarray}
Setting $\beta =\beta _{\alpha \alpha }=0$, we find the nontrivial stable
fixed point value
\begin{equation}
u_{\ast }=\frac{2\varepsilon }{3}\Biggl(1+\biggl(\frac{169}{288}+\frac{53}{
144}\ln \frac{4}{3}\biggr)\varepsilon \Biggr)+O(\varepsilon ^{2})  \label{41}
\end{equation}
from which the scaling exponents Eqs.\ (\ref{23},\ref{25}) of the Green
functions Eq.\ (\ref{27}) follow to second order of the $\varepsilon $
-expansion as
\begin{eqnarray}
\eta &=&-\frac{\varepsilon }{6}\Biggl(1+\biggl(\frac{25}{288}+\frac{161}{144}
\ln \frac{4}{3}\biggr)\varepsilon +O(\varepsilon ^{2})\Biggr)\ ,  \nonumber
\\
z &=&2-\frac{\varepsilon }{12}\Biggl(1+\biggl(\frac{67}{288}+\frac{59}{144}
\ln \frac{4}{3}\biggr)\varepsilon +O(\varepsilon ^{2})\Biggr)\ ,  \nonumber
\\
\nu &=&\frac{1}{2}+\frac{\varepsilon }{16}\Biggl(1+\biggl(\frac{107}{288}-
\frac{17}{144}\ln \frac{4}{3}\biggr)\varepsilon +O(\varepsilon ^{2})\Biggr)\
.  \label{42}
\end{eqnarray}
The first two expansions have been known for a long time from Reggeon field
theory \cite{Bak74,BrDa74} where a different definition of the exponents is
used. The expansion of the exponent $\nu $ was presented by the author in
\cite{Ja81}. The order parameter exponent $\beta $, which enters the scaling
law for the mean particle number in the active state $<n>=M\sim |\tau
|^{\beta }$, follows from Eq.\ (\ref{27}) as
\begin{equation}
\beta =\nu \frac{d+\eta }{2}=1-\frac{\varepsilon
}{6}\Biggl(1-\biggl(\frac{ 11 }{288}-\frac{53}{144}\ln
\frac{4}{3}\biggr)\varepsilon +O(\varepsilon ^{2}) \Biggr)\ .  \label{43}
\end{equation}

We now turn to the interspecies coupling constants $u_{\alpha \beta }$
with $ \alpha \neq \beta $. The fixed point values as the solutions of the
equation $\beta _{\alpha \beta }=0$, where $\beta _{\alpha \beta }$ is the
Gell-Mann--Low function Eq.\ (\ref{40a}), are of three different types:

\begin{enumerate}
\item  the decoupled fixed point $u_{12*}=u_{21*}=0$, totally instable for $
u=u_{*}$;\

\item  the two unidirectional coupled fixed points $u_{12*}=0$, $
u_{21*}=2u_{*}+O(\varepsilon ^{3})$ and $u_{21*}=0$, $u_{12*}=2u_{*}+O(
\varepsilon ^{3})$;

\item  the symmetric fixed point $u_{12*}=u_{21*}=u_{*}+O(\varepsilon ^{3})$.
\end{enumerate}

To discuss the stability and the crossover between the last two types we
try an Ansatz of the form $u_{\alpha \beta }=w+v\epsilon _{\alpha \beta }$
with $ \epsilon _{12}=-\epsilon _{21}=1$. We already know from the
one-loop result that $u_{\alpha \beta }$ is driven by the renormalization
flow to the fixed line $w=u_{\ast }$ with a crossover exponent $\phi
_{w}=\varepsilon /3+O(\varepsilon ^{2})$. Setting therefore $w=u$ in
$\beta _{\alpha \beta }$ we get
\begin{eqnarray}
\beta _{\alpha \beta }(u,v) &=&\beta (u)+\varepsilon _{\alpha \beta }\beta
_{v}(u,v)\ ,  \nonumber \\ \beta _{v}(u,v)
&=&-\biggl(\frac{1}{8}-\frac{3}{8}\ln \frac{4}{3}\biggr)
\Bigl(u^{2}-v^{2}\Bigr)v=-0.017\Bigl(u^{2}-v^{2}\Bigr)v\ .  \label{44}
\end{eqnarray}
For $u=u_{\ast }$ this equation shows that the symmetric fixed point $
v_{\ast }=0$ is instable in contrast to the stable unidirectional coupled
fixed points $v_{\ast }=\pm u_{\ast }$. The solution of the flow equation
$ ld \bar{v}/dl=\beta _{v}(u_{\ast },\bar{v})$ leads to the crossover
\begin{equation}
\bar{v}(l)^{2}=\frac{u_{\ast }^{\,2}}{1+\Bigl(u_{\ast
}^{\,2}/v^{2}-1\Bigr)l^{\phi _{v}}}  \label{45}
\end{equation}
with the very small crossover exponent
\begin{equation}
\phi _{v}=\biggl(\frac{1}{4}-\frac{3}{4}\ln \frac{4}{3}\biggr)u_{\ast
}^{\,2}=\biggl(\frac{1}{9}-\frac{1}{3}\ln \frac{4}{3}\biggr)\varepsilon
_{\ast }^{\,2}=0.0152\,\varepsilon ^{2}\ .  \label{46}
\end{equation}
A qualitative picture of the flow of the interspecies couplings in the
plane $u=u_{\ast }$ under renormalization is shown in FIG.~4.

\begin{center}
\epsfig{file=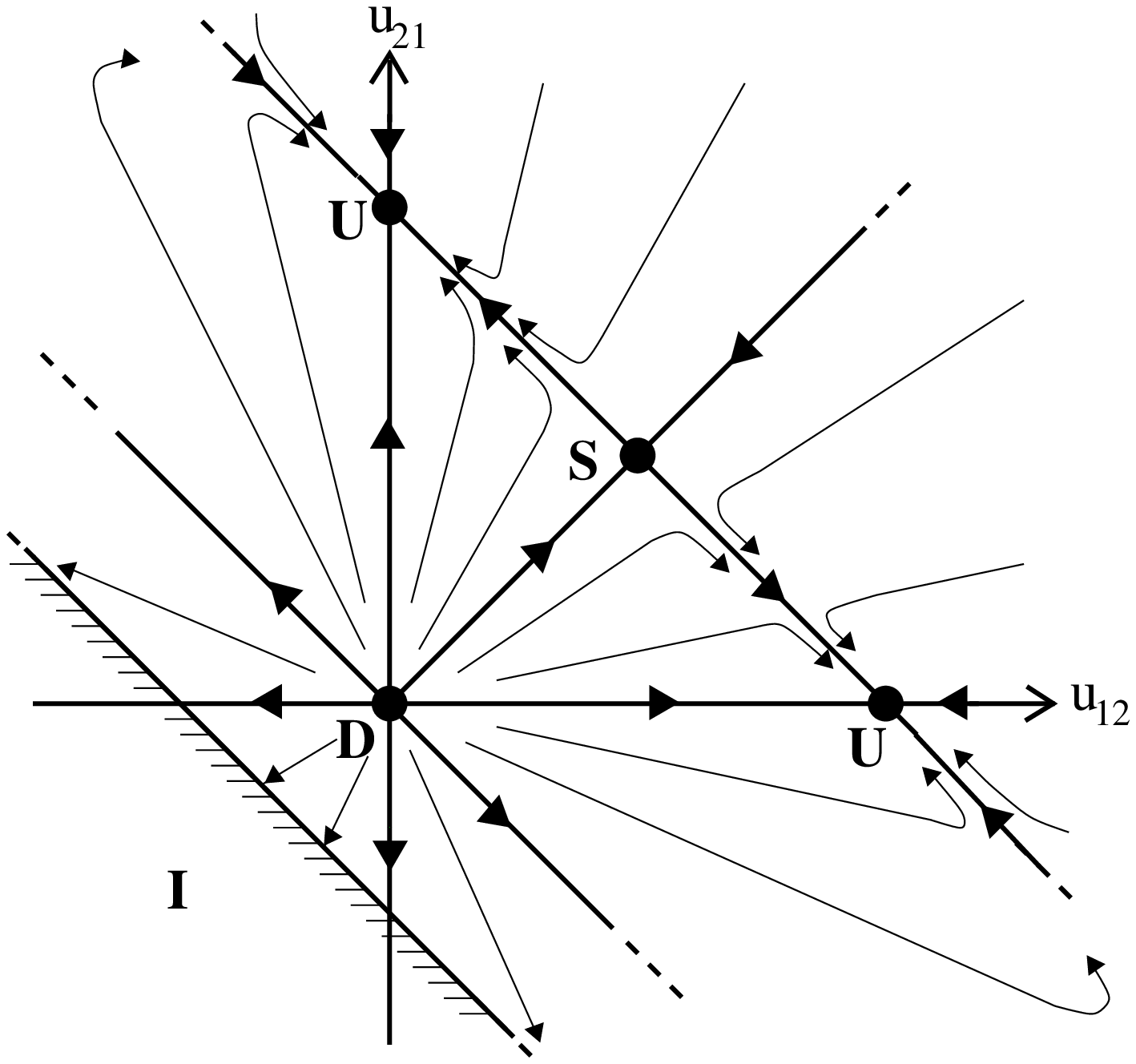,width=10cm}

FIG.\ 4. Flow of the interspecies couplings under renormalization
\end{center}

In addition to the four fixed points D (decoupled), S (symmetric), U
(unidirectional), the topology of the flow is determined by the symmetry
line $u_{12}=u_{21}$ which acts in the first quadrant as a separatrix
between the regions of attraction of the two unidirectional fixed points.
There exists another separatrix at the border of these regions, given to
first order in $\varepsilon $ by $u_{12}+u_{21}=0$, where the flow is
driven to the hatched line $u_{12}+u_{21}=-2u_{*}$. On the left of this
line we have the region of instability I, in which the condition
$\sum_{\alpha ,\beta }u_{\alpha \beta }n_{\alpha }n_{\beta }\geq 0$ for
all positive $ n_{\alpha }$ is violated. Therefore we conjecture that
interspecies couplings with $u_{12}+u_{21}<0$ ultimately lead to first
order transitions. In FIG.~4 the fixed point line of the one-loop
calculation is also shown. This line is the support of the slow crossover
to the unidirectional fixed points which therefore describe the ultimate
critical behavior of the MDP universality class.

We conjecture that in the case of colors with different flavors, $\lambda
_{\alpha }\neq \lambda _{\beta }$, all properties are smooth functions of
the ratio $\lambda _{\alpha }/\lambda _{\beta }$ as long as this ratio is
sufficiently close to one. Thus generalizing to different flavors, the
topology of the renormalization flow displayed in FIG.~4 should only be
smoothly deformed but not destroyed. In particular, the unidirectional
coupled fixed points U should be stable also for $\lambda _{\alpha }\neq
\lambda _{\beta }$.

\section{Symmetries and General Fixed Point Properties}

Having found the detailed fixed point structure of the model Eq.\
(\ref{33}) of two colors with the same flavor in the two-loop
approximation, we will now investigate which results are valid to all
orders of perturbation theory. According to the considerations of the
third chapter, one demonstrates easily that for $\alpha \neq \beta $ the
vertex function $ \Gamma _{\alpha ,\alpha \beta }=0$ if $g_{\alpha \beta
}=0$. Thus the lines of unidirectionally coupled models in FIG.~4, namely
$u_{12}=0$ or $u_{21}=0$ , respectively, are invariant under the
renormalization flow. Trivially, the decoupled fixed point D is the
intersection point of these lines. For $ g_{12}=g_{21}$ the dynamic
functional ${\cal J}$ Eq.\ (\ref{33}) possesses the symmetry
$s_{1}\leftrightarrow s_{2}$, $\widetilde{s}_{1}\leftrightarrow
\widetilde{s}_{1}$, and $\tau _{1}\leftrightarrow \tau _{2}$ from which we
find that $\Gamma _{1,12}=\Gamma _{2,21}$. Thus, these two vertex
functions need the same $Z$-factor: $Z_{12}=Z_{21}$. It follows that the
symmetry line in FIG.~4, $u_{12}=u_{21}$, is invariant under
renormalization.

Is there a condition that determines the crossover line? The answer is yes.
We change to variables corresponding to the total and relative particle
numbers, respectively,
\begin{eqnarray}
s &=&s_{1}+s_{2\ ,\qquad }\widetilde{s}=\frac{1}{2}\Bigl(\widetilde{s}_{1}+
\widetilde{s}_{2}\Bigr)\ ,  \nonumber \\
c &=&s_{1}-s_{2\ ,\qquad }\widetilde{c}=\frac{1}{2}\Bigl(\widetilde{s}_{1}-
\widetilde{s}_{2}\Bigr)\ .  \label{47}
\end{eqnarray}
Such linear transformations do not alter the measure of the functional
integrals, and the dynamic functional changes, in the special case $\tau
_{1}=\tau _{2}=\tau $, to
\begin{eqnarray}
{\cal J} &=&\int dtd^{d}x\,\lambda \Biggl(\widetilde{s}\biggl(\lambda
^{-1}\partial _{t}+\bigl(\tau -\nabla ^{2}\bigr)+\Bigl(\frac{g}{4}+\frac{
g_{12}+g_{21}}{8}\Bigr)s-\frac{g}{2}\widetilde{s}\biggr)s  \nonumber \\
&&+\widetilde{c}\biggl(\lambda ^{-1}\partial _{t}+\bigl(\tau -\nabla ^{2}
\bigr)+\frac{g}{2}s\biggr)c-\frac{g}{2}\widetilde{c}^{2}s-g\widetilde{s}
\widetilde{c}c  \nonumber \\
&&+\biggl(\frac{g}{4}-\frac{g_{12}+g_{21}}{8}\biggr)\widetilde{s}c^{2}+\frac{
g_{12}-g_{21}}{8}\widetilde{c}\biggl(s^{2}-c^{2}\biggr)\Biggr)\ .
\label{48}
\end{eqnarray}
We see that in the case $g_{12}+g_{21}=2g$ the dynamics of the total
particle number decouples from the dynamics of the relative one in the
sense that all vertex functions containing $\widetilde{s}$-, but no
$\widetilde{c}$ -legs are zero if $c$-legs are attached. In particular
$\Gamma _{\tilde{s} ,cc}=0$ and $\Gamma _{\tilde{s},ss}=-\Gamma
_{\tilde{s}\tilde{s},s}$, which leads to the renormalized interspecies
couplings with $u_{12}+u_{21}=2u$ and determines the crossover line to all
loop-orders. Note that on the symmetry line $g_{12}=g_{21}$ holds, and the
functional ${\cal J}$ is then invariant against $c\leftrightarrow -c$,
$\widetilde{c} \leftrightarrow -\widetilde{c}$ .

The different fixed point values of the interspecies couplings are now
fully determined as the intersection points of the several invariant lines
(see FIG.\ 4): the unidirectional lines $u_{12}=0$, $u_{21}=0$, the
symmetry line $u_{12}=u_{21}$, and the crossover line $u_{12}+u_{21}=2u$.
Thus, we find to all orders for the symmetric fixed point S: $u_{12\ast
}=u_{21\ast }=u_{\ast }$ and for the stable unidirectional fixed points U:
$u_{12\ast }=0,$ $ u_{21\ast }=2u_{\ast }$ and $u_{21\ast }=0,$ $u_{12\ast
}=2u_{\ast }$, respectively. These are of course the results that we have
found explicitly in the two-loop approximation.

The unidirectionally coupled model, which describes the generic scaling
properties of the MDP processes, exhibits another symmetry. Using
$g_{12}=0$ and $g_{21}=2g^{\prime }$, we write the dynamic functional of
this model in the form
\begin{eqnarray}
{\cal J}_{u} &=&\int dtd^{d}x\,\lambda \Biggl(\widetilde{s}_{1}\biggl(
\lambda ^{-1}\partial _{t}+\bigl(\tau _{1}-\nabla ^{2}\bigr)+\frac{g}{2}
\bigl(s_{1}-\widetilde{s}_{1}\bigr)\biggr)s_{1}  \nonumber \\
&&+\widetilde{s}_{2}\biggl(\lambda ^{-1}\partial _{t}+\bigl(\tau
_{2}-\nabla
^{2}\bigr)+\frac{g}{2}\bigl(s_{2}-\widetilde{s}_{2}\bigr)\biggr)s_{2}+
\widetilde{s}_{2}\biggl(-\sigma +g^{\prime }s_{2}\biggr)s_{1}\Biggr)\ .
\label{49}
\end{eqnarray}
Here we have introduced a further harmonic unidirectional coupling $\sim
\sigma $, which corresponds to an additional linear term: $\partial
n_{2}/\partial t=\cdots +\lambda \sigma n_{1}$ in the Langevin equation
for species $2$. This term was first considered by T\"{a}uber et al.
\cite{THH98} in their study of the nonequilibrium critical behavior in
unidirectionally coupled DP processes. Such a term does not alter the
general renormalizations as long as the corresponding composed field
$\widetilde{s} _{2}s_{1}$ is treated as a soft insertion. Of course,
$\sigma $ is a relevant parameter like the temperatures $\tau _{\alpha }$
and needs its own renormalization factor $Z_{\sigma }$ determined by
\begin{equation}
\sigma \rightarrow \mathaccent"7017\sigma =Z_{\sigma }Z_{\lambda
}^{\,-1}\sigma \ ,  \label{50}
\end{equation}
in such a way that the renormalized vertex function with an insertion $
\Gamma _{2,1;(\widetilde{s}_{2}s_{1})}$ is finite. The settings $\sigma
=\tau _{1}=\tau _{2}=0$ define the multicritical point.

From Reggeon field theory one knows a transformation called rapidity
reversal: $s(t)\leftrightarrow -\widetilde{s}(-t)$, which is broken by the
DP transition to an active state. Here we generalize it to read
\begin{eqnarray}
s_{1}(t) &\rightarrow &-\widetilde{s}_{2}(-t)\ ,\qquad \widetilde{s}
_{1}(t)\rightarrow -s_{2}(-t)-s_{1}(-t)\ ,  \nonumber \\
\widetilde{s}_{2}(t) &\rightarrow &-s_{1}(-t)\ ,\qquad s_{2}(t)\rightarrow
- \widetilde{s}_{1}(-t)+\widetilde{s}_{2}(-t)\ .  \label{51}
\end{eqnarray}
Under this transformation the functional ${\cal J}_{u}$ changes to
\begin{eqnarray}
{\cal J}_{u} &\rightarrow &\int dtd^{d}x\,\lambda \Biggl(\widetilde{s}_{1}
\biggl(\lambda ^{-1}\partial _{t}+\bigl(\tau _{2}-\nabla
^{2}\bigr)+\frac{g}{ 2}\bigl(s_{1}-\widetilde{s}_{1}\bigr)\biggr)s_{1}
\nonumber \\ &&+\widetilde{s}_{2}\biggl(\lambda ^{-1}\partial
_{t}+\bigl(\tau _{1}-\nabla
^{2}\bigr)+\frac{g}{2}\bigl(s_{2}-\widetilde{s}_{2}\bigr)\biggr)s_{2}
\nonumber \\ &&+\widetilde{s}_{2}\biggl(-\sigma +\tau _{1}-\tau
_{2}+g^{\prime }s_{2} \biggr)s_{1}+\bigl(g-g^{\prime
}\bigr)\biggl(\widetilde{s}_{1}\widetilde{s}
_{2}s_{1}-\widetilde{s}_{2}^{\,2}s_{1}+\widetilde{s}_{2}s_{1}s_{2}\biggr)
\Biggr)\ .  \label{52}
\end{eqnarray}
We learn from this relation that in the case $g=g^{\prime }$, ${\cal
J}_{u}$ gains a higher symmetry at the multicritical point, which is not
destroyed by renormalization. It follows in this case that $\sigma $
renormalizes in the same way as the $\tau _{\alpha }$, which implies
$Z_{\sigma }=Z_{\tau }$ , and that the renormalized couplings are related
by $u_{*}^{\prime }=u_{21*}=u_{*}$ at the fixed point, as we already know
from above. From these considerations follows that the crossover exponent,
which is defined by the scaling invariants $\sigma /\tau _{\alpha
}^{\,\Phi }$, is given by $ \Phi =1$.

\section{The THHG Model}

Recently T\"{a}uber et al.\ \cite{THH98} (in the following abbreviated by
THHG) introduced a general unidirectionally coupled DP process with
species-independent diffusion coefficients, that was motivated by a study of
Alon et al.\ \cite{AEHM96} on a nonequilibrium growth model for adsorption
and desorption of particles which displays a roughening transition.

The THHG-model reads in our dynamic functional language as
\begin{eqnarray}
{\cal J}_{THHG} &=&\int dtd^{d}x\lambda \Biggl(\widetilde{s}_{1}\biggl(
\lambda ^{-1}\partial _{t}+\tau _{1}-\nabla ^{2}+\frac{g}{2}\bigl(s_{1}-
\widetilde{s}_{1}\bigr)\biggr)s_{1}  \nonumber \\
&&+\widetilde{s}_{2}\biggl(\lambda ^{-1}\partial _{t}+\tau _{2}-\nabla
^{2}+ \frac{g}{2}\bigl(s_{2}-\widetilde{s}_{2}\bigr)\biggr)s_{2}
\nonumber \\ &&-\widetilde{s}_{2}\biggl(\sigma +b\nabla
^{2}-\frac{f_{1}}{2} s_{1}-f_{2}s_{2}+f_{2}^{\prime
}\widetilde{s}_{1}+\frac{f_{1}^{\prime }}{2}
\widetilde{s}_{2}\biggr)s_{1}\Biggr)\ .  \label{53}
\end{eqnarray}
In contrast to THHG, we have introduced an additional cross diffusion term
$ \sim \widetilde{s}_{2}\nabla ^{2}s_{1}$ here that is indispensable for a
complete renormalization of the general model. In physical terms, coarse
graining will always produce cross diffusion in this coupled model. But
coarse graining does more: it also produces a term proportional to the
time derivative of the density of the first species in the Langevin
equation of the second one -- besides further irrelevant couplings.
Accordingly, in the functional ${\cal J}_{THHG}$ a term proportional to
$\widetilde{s} _{2}\partial _{t}s_{1}$ arises. However, such a term can be
eliminated by a suitable redefinition of the fields
\begin{equation}
\widetilde{s}_{1}\rightarrow \widetilde{s}_{1}-\beta \widetilde{s}_{2}\
,\quad s_{1}\rightarrow s_{1}\ ,\qquad \widetilde{s}_{2}\rightarrow
\widetilde{s}_{2}\ ,\quad s_{2}\rightarrow s_{2}-\beta s_{1}\ ,  \label{54}
\end{equation}
so that the harmonic parts with the time derivatives in the dynamic
functional remain diagonal. Note that in the special case $
b=f_{1}=f_{1}^{\prime }=f_{2}^{\prime }=0$, $f_{2}=g^{\prime }$ the
functional ${\cal J}_{THHG}$ Eq.~(\ref{53}) is identical to ${\cal
J}_{u}$, Eq.~(\ref{49}), from which we know that it is fully
renormalizable, and, in particular, in the case $g^{\prime }=g$ by the
$Z$-factors Eqs.\ (\ref{36},\ref{39}). In the following we will
demonstrate that the asymptotic properties of the THHG-model indeed belong
to the universality class described by ${\cal J}_{u}$.

It is well known \cite{We74} that the infinitesimal generators of a
continuous transformation of the fundamental fields, which leads to a
forminvariance of the describing statistical functional, define redundant
operators (composite fields). In the case that these redundant operators
are relevant or marginal, they unnecessarily contaminate the
renormalization group. Therefore they should be avoided from the outset.
Here we introduce a linear, homogeneous transformation between the fields,
which does not change the form of the functional ${\cal J}_{THHG}$ Eq.\
(\ref{53}). Let us call it the $\alpha $-transformation:
\begin{equation}
\widetilde{s}_{1}\rightarrow \widetilde{s}_{1}-\alpha \widetilde{s}_{2}\
,\quad s_{1}\rightarrow s_{1}\ ,\qquad \widetilde{s}_{2}\rightarrow
\widetilde{s}_{2}\ ,\quad s_{2}\rightarrow s_{2}+\alpha s_{1}\ .  \label{55}
\end{equation}
Note the difference to the $\beta $-transformation Eq.\ (\ref{54}), which
is exploited to eliminate a $\widetilde{s}_{2}\partial _{t}s_{1}$-term
from $ {\cal J}_{THHG}$. This $\alpha $-transformation leads to new
coupling constants that we denote with a bar:
\begin{eqnarray}
\bar{\sigma} &=&\sigma +\alpha \left( \tau _{1}-\tau _{2}\right) \ ,
\nonumber \\
\bar{f}_{1} &=&f_{1}+2\alpha f_{2}+\alpha \left( \alpha -1\right) g\ ,\qquad
\bar{f}_{2}=f_{2}+\alpha g\ ,  \nonumber \\
\bar{f}_{1}^{\prime } &=&f_{1}^{\prime }-2\alpha f_{2}^{\prime }+\alpha
\left( \alpha +1\right) g\ ,\qquad \bar{f}_{2}^{\prime }=f_{2}^{\prime
}-\alpha g\ .  \label{56}
\end{eqnarray}

Now we renormalize the THHG-model by the scheme
\begin{eqnarray}
\widetilde{s}_{1} &\rightarrow &\mathaccent"7017{\widetilde{s}}
_{1}=Z_{s}^{\,1/2}\bigl(\widetilde{s}_{1}+A\widetilde{s}_{2}\bigr)\;,\qquad
s_{1}\rightarrow \mathaccent"7017{s}_{1}=Z_{s}^{\,1/2}s_{1}\;,  \nonumber
\\ s_{2} &\rightarrow
&\mathaccent"7017s_{2}=Z_{s}^{\,1/2}\bigl(s_{2}+As_{1} \bigr)\;,\qquad
\widetilde{s}_{2}\rightarrow \mathaccent"7017{\widetilde{s}}
_{2}=Z_{s}^{\,1/2}\widetilde{s}_{2}\;,  \nonumber \\ \tau _{\alpha }
&\rightarrow &\mathaccent"7017\tau _{\alpha }=Z_{\lambda }^{\,-1}Z_{\tau
}\tau _{\alpha }\ ,\qquad g\rightarrow \mathaccent
"7017g=Z_{s}^{\,-1/2}Z_{\lambda }^{\,-1}Z_{g}g\ ,  \nonumber \\ \sigma
&\rightarrow &\mathaccent"7017\sigma =Z_{\lambda }^{\,-1}\Bigl(Z_{\sigma
}\sigma +\bigl(Y_{1}+AZ_{\tau }\bigr)\tau _{1}+ \bigl( Y_{2}+AZ_{\tau
}\bigr)\tau _{2}\Bigr)\;,  \nonumber \\ b &\rightarrow
&\mathaccent"7017b=Z_{\lambda }^{\,-1}Z_{b}b-2A\;,  \nonumber
\\
f_{1} &\rightarrow &\mathaccent"7017f_{1}=Z_{s}^{\,-1/2}Z_{\lambda
}^{\,-1}\Bigl(Z_{1}f_{1}-A\left( 1-A\right) Z_{g}^{\,}g-2AZ_{2}f_{2}\Bigr)\;,
\nonumber \\
f_{1}^{\prime } &\rightarrow &\mathaccent"7017f_{1}^{\prime
}=Z_{s}^{\,-1/2}Z_{\lambda }^{\,-1}\Bigl(Z_{1}^{\prime }f_{1}^{\prime
}-A\left( 1-A\right) Z_{g}^{\,}g-2AZ_{2}^{\prime }f_{2}^{\prime }\Bigr)\;,
\nonumber \\
f_{2} &\rightarrow &\mathaccent"7017f_{2}=Z_{s}^{\,-1/2}Z_{\lambda
}^{\,-1}\Bigl(Z_{2}f_{2}-AZ_{g}^{\,}g\Bigr)\;,  \nonumber \\
f_{2}^{\prime } &\rightarrow &\mathaccent"7017f_{2}^{\prime
}=Z_{s}^{\,-1/2}Z_{\lambda }^{\,-1}\Bigl(Z_{2}^{\prime }f_{2}^{\prime
}-AZ_{g}^{\,}g\Bigr)\;.  \label{57}
\end{eqnarray}
where $g=(u\mu ^{\varepsilon }/G_{\varepsilon })^{1/2}$, $
Z_{g}=Z_{u}^{\,1/2} $, and the $Z_{i}$ with $i=s,\lambda ,\tau ,u$ are
given by Eqs.\ (\ref{36},\ref{39}). Besides $u$, dimensionless coupling
constants are defined as $v_{\alpha }^{(^{\prime })}=(G_{\varepsilon }/\mu
^{\varepsilon })^{1/2}f_{\alpha }^{(^{\prime })}$. The scheme Eq.\
(\ref{57}) is chosen in such a way that the renormalized dynamic
functional reads
\begin{eqnarray}
{\cal J}_{THHG} &=&\int dtd^{d}x\lambda \Biggl(\widetilde{s}_{1}\biggl(
Z_{s}\lambda ^{-1}\partial _{t}+Z_{\tau }\tau _{1}-Z_{\lambda }\nabla
^{2}+ \frac{Z_{g}g}{2}\bigl(s_{1}-\widetilde{s}_{1}\bigr)\biggr)s_{1}
\nonumber \\ &&+\widetilde{s}_{2}\biggl(Z_{s}\lambda ^{-1}\partial
_{t}+Z_{\tau }\tau _{2}-Z_{\lambda }\nabla
^{2}+\frac{Z_{g}g}{2}\bigl(s_{2}-\widetilde{s}_{2} \bigr)\biggr)s_{2}
\nonumber \\ &&+\widetilde{s}_{2}\biggl(2AZ_{s}\lambda ^{-1}\partial
_{t}-Z_{\sigma }\sigma -Y_{1}\tau _{1}-Y_{2}\tau _{2}-Z_{b}b\nabla ^{2}
\nonumber \\ &&+\frac{Z_{1}f_{1}}{2}s_{1}+Z_{2}f_{2}s_{2}-Z_{2}^{\prime
}f_{2}^{\prime } \widetilde{s}_{1}-\frac{Z_{1}^{\prime }f_{1}^{\prime
}}{2}\widetilde{s}_{2} \biggr)s_{1}\Biggr)\ .  \label{58}
\end{eqnarray}
Note that the counter term $2AZ_{s}$ serves to cancel primitive divergencies
arising in the vertex function $\partial_{\omega}\Gamma_{{\widetilde{s}}_2
s_1}$. In dimensional regularization and minimal renormalization the
counterterms $A $ and $Y_{\alpha }$ are given by series in $\varepsilon
^{-1} $ beginning with simple poles $\sim A^{(1)} /\varepsilon $ and $\sim
Y_{\alpha }^{(1)}/\varepsilon $, respectively. We have formerly shown that
only the residua of these poles determine the renormalization group
functions.

It is now appropriate to define $\alpha $-transformation invariant
dimensionless coupling constants as
\begin{equation}
w_{1}=\sqrt{u}\left( v_{1}+v_{2}\right) -v_{2}^{\,2}\;,\quad w_{1}^{\prime
}= \sqrt{u}\left( v_{1}^{\prime }+v_{2}^{\prime }\right) -v_{2}^{\prime
\,2}\;,\quad w_{2}=\sqrt{u}\left( v_{2}+v_{2}^{\prime }\right) \;.
\label{59}
\end{equation}

The somewhat lengthy but simple calculation of all the one-loop
renormalizations leads to the Gell-Mann--Low functions of the
renormalization group equation. In part, in the case $b=0$, they can be
derived from the results of THHG \cite{THH98}. If we set $u$ to its fixed
point value $u_{*}$ and define as usual $\beta _{p}=\left. \partial
p/\partial \ln \mu \right| _{0}$, where $p$ is any of the coupling
constants, we obtain for the $\beta $ -functions of the invariant couplings
\begin{eqnarray}
8\beta _{w_{1}}
&=&4\bigl(2w_{1}+w_{2}-u_{*}\bigr)w_{2}+4u_{*}w_{1}^{\prime
}+16u_{*}a-u_{*}b\bigl(8w_{1}+10w_{2}\bigr)+9\bigl( u_{*}b\bigr)^{2}\;,
\nonumber \\ 8\beta _{w_{1}^{\prime }} &=&4\bigl(2w_{1}^{\prime
}+w_{2}-u_{*}\bigr) w_{2}+4u_{*}w_{1}+16u_{*}a-u_{*}b\bigl(8w_{1}^{\prime
}+10w_{2}\bigr)+9\bigl( u_{*}b\bigr)^{2}\;,  \nonumber \\ 4\beta _{w_{2}}
&=&8\bigl(w_{2}-u_{*}\bigr)w_{2}+4u_{*}\bigl( w_{1}+w_{1}^{\prime
}\bigr)+8u_{*}a-6u_{*}bw_{2}+3\bigl(u_{*}b\bigr)^{2}\;, \nonumber \\
8u_{*}\beta _{b} &=&u_{*}\bigl(w_{1}+w_{1}^{\prime }\bigr)+\bigl(
w_{2}-u_{*} \bigr)w_{2}+16u_{*}a-u_{*}b\bigl(w_{2}+u_{*}\bigr)
+\bigl(u_{*}b\bigr)^{2}\;. \label{63}
\end{eqnarray}
The function $a$ results from the additive renormalization $A$ and mixes the
renormalized fields in a correlation or response function under application
of the renormalization group as
\begin{equation}
\biggl[{\cal D}_{\mu }+\frac{\gamma }{2}\biggr]\bigl\{\tilde{s}_{1},s_{1},
\tilde{s}_{2},s_{2}\bigr\}=\bigl\{-a\tilde{s}_{2},0,0,-as_{1}\bigr\}\;.
\label{64}
\end{equation}
Here ${\cal D}_{\mu }$ is the renormalization group differential operator
now given by
\begin{equation}
{\cal D}_{\mu }=\mu \partial _{\mu }+\zeta \lambda \partial _{\lambda
}+\kappa _{\tau }\tau \partial _{\tau }+\bigl(\kappa _{\sigma }\sigma
+\kappa _{1}\tau _{1}+\kappa _{2}\tau _{2}\bigr)\partial _{\sigma
}+\sum_{p}\beta _{p}\partial _{p}\;,  \label{65}
\end{equation}
with $\kappa _{\sigma }=\gamma _{\lambda }-\gamma _{\sigma },$ $\kappa
_{\alpha }=-y_{\alpha }-a$, and Eq.\ (\ref{64}) acts on Green functions.
The $y_{\alpha }=-\sum_{p}\partial _{p}Y_{\alpha }^{(1)}$ result from the
additive renormalizations $Y_{\alpha }$. The function $a=-\sum_{p}\partial
_{p}A^{(1)}$ is found to be
\begin{equation}
a=-\frac{1}{16}\biggl(2w_{1}+w_{1}^{\prime }+2\Bigl(\frac{w_{2}}{u}-1\Bigr
)w_{2}-4bw_{2}+3ub^{2}\biggr)\;.  \label{66}
\end{equation}
The new renormalizations yield
\begin{eqnarray}
\gamma _{\sigma } &=&\frac{1}{2}\Bigl(bu-w_{2}\Bigr)\;,  \nonumber \\
y_{1} &=&\frac{1}{2}\biggl(\sqrt{u}v_{1}+v_{2}v_{2}^{\prime }-\frac{3b}{2}
\sqrt{u}v_{2}-\frac{b}{2}\sqrt{u}v_{2}^{\prime
}+\frac{3b^{2}}{4}u\biggr)\;, \nonumber \\ y_{2}
&=&\frac{1}{2}\biggl(\sqrt{u}v_{1}^{\prime }+v_{2}v_{2}^{\prime }-
\frac{3b}{2}\sqrt{u}v_{2}^{\prime
}-\frac{b}{2}\sqrt{u}v_{2}+\frac{3b^{2}}{4} u\biggr)\;.  \label{67}
\end{eqnarray}

In order to determine the fixed-point solutions of Eqs.\ (\ref{63}),
$\beta _{p*}=0$, we impose the condition $a_{*}=b_{*}=0$. This yields $
w_{1*}=w_{1*}^{\prime }=0$, with $w_{2*}=0$ (unstable) or $w_{2*}=u_{*}$
(stable). It can easily be checked that these solutions are consistent
with the full set of Eqs.\ (\ref{63}) and (\ref{66}). These are the
solutions found by THHG \cite{THH98}. Note that on the fixed point lines
generated from the stable fixed point by the $\alpha$-transformation a
minimally coupled fixed point with $v_{1*}=v_{1*}^{\prime }=v_{2*}^{\prime
}=0,\;v_{2*}=\sqrt{u_{*}}$ is found.

Now one has to prove stability of the fixed points of the full equations
(\ref{63}) without using the constraints $a_{*}=b_{*}=0$. A linearization
about $w_{1}=w_{1}^{\prime }=b=0$ and either $w_{2}=0$ or $w_{2}=u_{*}$
shows that the flow of $b$ is unstable for $w_{2}=0$, whereas it shows
full stability of the fixed point for $w_{2}=u^{*}$. This vindicates the
neglect of $a,b$ and the corresponding counterterms $A$ and $Z_{b}b$ in
\cite{THH98} but only at the stable fixed point line, which is generated
from the fixed point by the $\alpha $-transformation. However without
further knowledge this statement is only correct in the one-loop
calculation and could be violated in higher loop orders. We will show that
the stable fixed point is given by $w_{1*}=w_{1*}^{\prime }=0$,
$w_{2*}=u_{*}$ to all orders of the loop expansion. As a consequence, the
fixed point values $ a_{*},b_{*},y_{1*},y_{2*}$ are zero, and $\gamma
_{\sigma *}=\gamma _{\tau *} $. This leads to a crossover exponent $\Phi
=1$ where $\Phi $ determines the scaling of $\sigma /|\tau _{i}|^{\Phi }$.

Indeed, we see from Eq.\ (\ref{59}) that the stable one-loop order fixed
point belongs, up to an $\alpha $-transformation, to the dynamic
functional $ {\cal J}_{THHG}$, Eq.\ (\ref{58}), with coupling constants $
f_{1}=f_{1}^{\prime }=f_{2}^{\prime }=0$ and $f_{2}=g$, i.e.\ model ${\cal
J} _{u}$, Eq.~(\ref{53}), with the additional constraint $g^{\prime }=g$.
Above it was shown that this equality leads to rapidity reversal Eq.\
(\ref{51}) as a higher symmetry. This higher symmetry is preserved under
renormalization and, because ${\cal J}_{u}$ is fully renormalizable, we
have the result $w_{1\ast }=w_{1\ast }^{\prime }=0$, $w_{2\ast }=u_{\ast
}$ and $ \Phi =1$ to all loop orders.

Computations based on the dynamic functional ${\cal J}_{u}$,
Eq.~(\ref{53}), are much easier to perform than calculations using the
complete model ${\cal J} _{THHG}$, Eq.~(\ref{58}). Thus it may be possible
to find the equation of state for $M_{2}=g\langle s_{2}\rangle $ to second
order and check the assumptions made in \cite{THH98} on the
reexponentiation of logarithms to yield the new order parameter exponent
$\beta _{2}$ of that paper (for a calculation of the equation of state for
$M_{1}=g\langle s_{1}\rangle $ to two-loop order see \cite{JKO98}).

The model ${\cal J}_{u}$ describes the coupled DP processes near the
multicritical point $\tau _{1}=\tau _{2}=\sigma =0$. What is needed for a
thorough calculation of $\beta _{2}$ is a theory that comprises the limit
$ \sigma \rightarrow \infty $. Therefore our considerations here do not
solve the problem addressed in \cite{THH98}, namely the determination of
the scaling exponent $\beta _{2}$ that controls the scaling $M_{2}\propto
\sigma ^{\beta _{1}-\beta _{2}}\left( \tau _{1c}-\tau _{1}\right) ^{\beta
_{2}}$ where species $1$ is in its active phase. THHG calculate $\beta
_{2}$ by reexponentiation of logarithms. (We have done a recalculation and
find a slightly different value $\beta _{2}=1/2-13\varepsilon
/96+O(\varepsilon ^{2})$. The difference arises from a subleading term
resulting from the ominous peculiar diagram FIG.~9(c) in \cite{THH98}.)
However the approach of THHG relies on the assumption that simple
reexponentiation is possible. To derive such scaling properties
faithfully, one indeed has to solve the crossover problem $\sigma
\rightarrow \infty $ which (possibly!) induces a new scaling at infinity
for the correlations of species $2$. Some features of this crossover
remind us of the crossover from special to ordinary behavior in the theory
of surface transitions \cite{Diehl}, with $\sigma $ corresponding to the
surface enhancement $c$ and the species $1$ and $2$ corresponding to the
bulk and surface respectively. The crossover problem of interest here is
thus as yet unsolved.

\section{Discussion}

In this paper, we have studied multicolored directed percolation processes
(MDP). In particular, we have shown that the scaling behavior of these
coupled DP processes near their absorbing state transition is determined
by the same critical exponents as known from simple one-species DP and
therefore is independent from the number of colors. The characteristic
asymptotic feature of MDP shows up in the asymptotic unidirectional
coupling of each pair of colors. A special result of our analysis is the
very slow crossover to this asymptotic unidirectionality which may be seen
in computer simulations. The unidirectional behavior of the couplings of
an interacting population is summarized in the following graphical
picture. Consider a graph where each node represents one color. The stable
fixed points are then represented by the so-called tournaments, that are
the complete graphs with directed edges \cite{Har69}. The directed edge
from color $\alpha $ to color $\beta $ stands for the influence of $\alpha
$ on $\beta $ in the respective equations of motion. In particular, for a
population of three species there exist two different tournaments (up to
permutation of the colors) which we call ``cyclic'' and ``hierarchic'',
FIG.~5.

\bigskip

\begin{center}
\epsfig{file=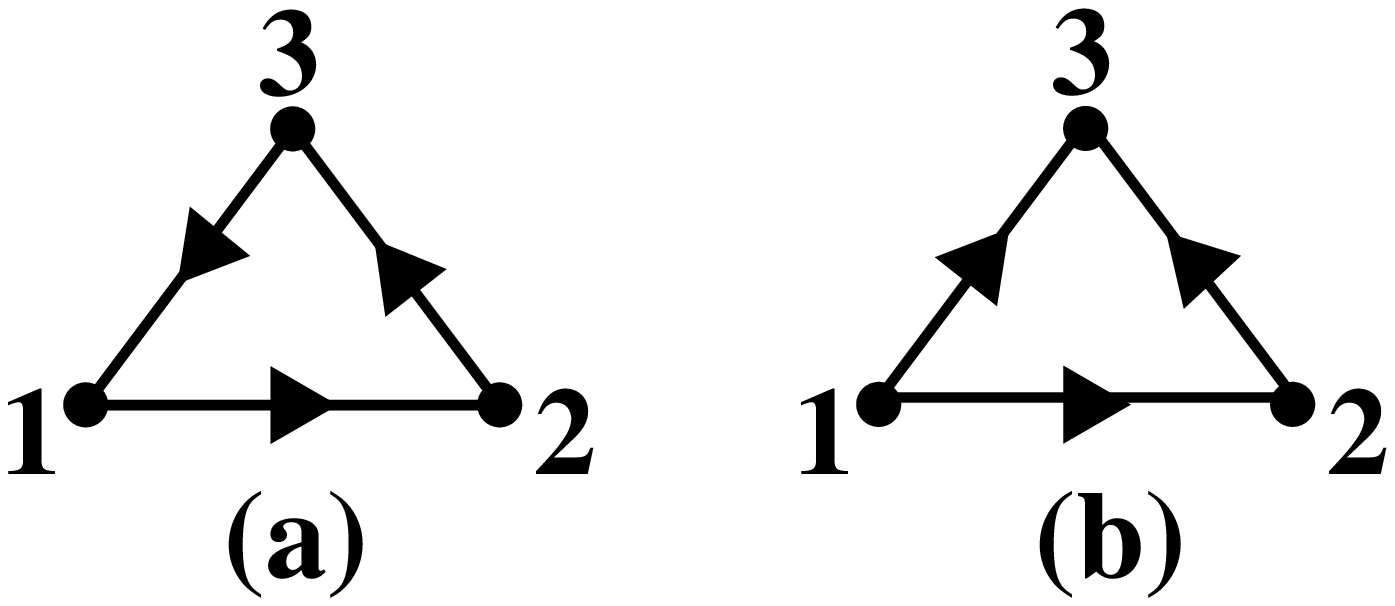,width=8cm}

FIG.\ 5. Cyclic and hierarchic tournament of three species
\end{center}

One gets a graphical picture for the not fully stable fixed points either by
deleting the directionality of the edges between the nodes for symmetric
couplings of the corresponding colors, or by completely deleting the edges
for the uncoupled pairs (uncomplete tournaments).

To get a simple qualitative impression of the behavior of the dynamic
system of a population corresponding to a tournament of colors with the
same flavor, we consider, in the active region with all$\,r_{\alpha
}:=-\tau _{\alpha }/2g\geq 0$, a renormalized mean-field theory of the
equations (\ref {1},\ref{2}) for spatially homogeneous densities
$\bar{n}_{a}$ and set $ g_{\alpha \beta }=g\left( 1+\varepsilon _{\alpha
\beta }\right) $ with $ \varepsilon _{\alpha \alpha }=0$. We redefine the
time scale $t\rightarrow 2t/\lambda g$ and get
\begin{equation}
\partial _{t}\bar{n}_{a}=\biggl(r_{\alpha }-\sum_{\beta }\bigl(1+\varepsilon
_{\alpha \beta }\bigr)\,\bar{n}_{\beta }\biggr)\,\bar{n}_{\alpha }\ .
\label{68}
\end{equation}
Then the decision, which of all the $\varepsilon _{\alpha \beta
}=-\varepsilon _{\beta \alpha }$ equal $\pm 1$, defines the tournament.
The edge between a pair of species is directed from $\beta $ to $\alpha $
if $ \varepsilon _{\alpha \beta }=-\varepsilon _{\beta \alpha }=+1$ and
vice versa. The directions of the edges therefore represent the
unidirectional ``pressure'' on the reproduction rate resulting from one
color to another. Despite the simplicity of the Eqs.\ (\ref{68}) they can
generate a complex dynamic behavior (see e.g.\ \cite{HoSi88,Mu89}). First,
let us consider the stationary states of Eqs.\ (\ref{68}) for a population
consisting of three species. In the ternary phase diagram spanned by the
positive rates $ r_{1},r_{2},r_{3}$ in the subspace $\sum_{\alpha
}r_{\alpha }=r=const.$, one finds different regions with one, two ore all
three species alive, FIG.~6.

\bigskip

\begin{center}
\epsfig{file=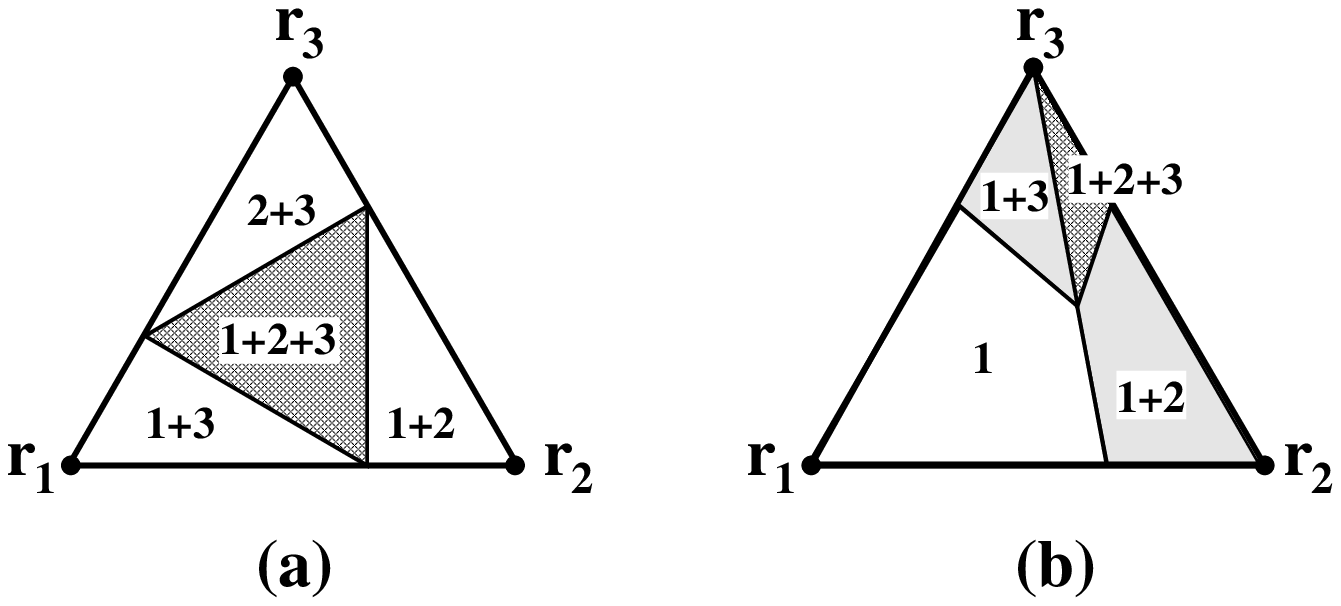,width=10cm}

FIG.\ 6. Phase diagrams of the cyclic and hierarchic tournament
\end{center}

These regions are bounded by critical lines with absorbing state
transitions where some colors become extinct. The dynamic behavior of the
hierarchic tournament FIG.~5(b) is relatively simple: from each
nonequilibrium initial state $\left\{ \bar{n}_{\alpha }^{(0)}\right\} $,
the system relaxes to the stationary state which is a stable node. In the
case of the cyclic tournament FIG.~5(a), for rates $\left\{ r_{\alpha
}\right\} $ such that we have a three species stationary state, one also
finds regions for which the ultimate relaxation behavior is characterized
by attracting nodes. But for rates which lead to stationary states
belonging to the crosshatched area in the ternary diagram FIG.~7 spanned
by the $\left\{ \bar{n}_{\alpha }\right\} $, we find stable spirals
(damped cyclic relaxation) as the ultimate relaxation behavior
(qualitatively pictured by the trajectory in FIG.~7).

\bigskip

\begin{center}
\epsfig{file=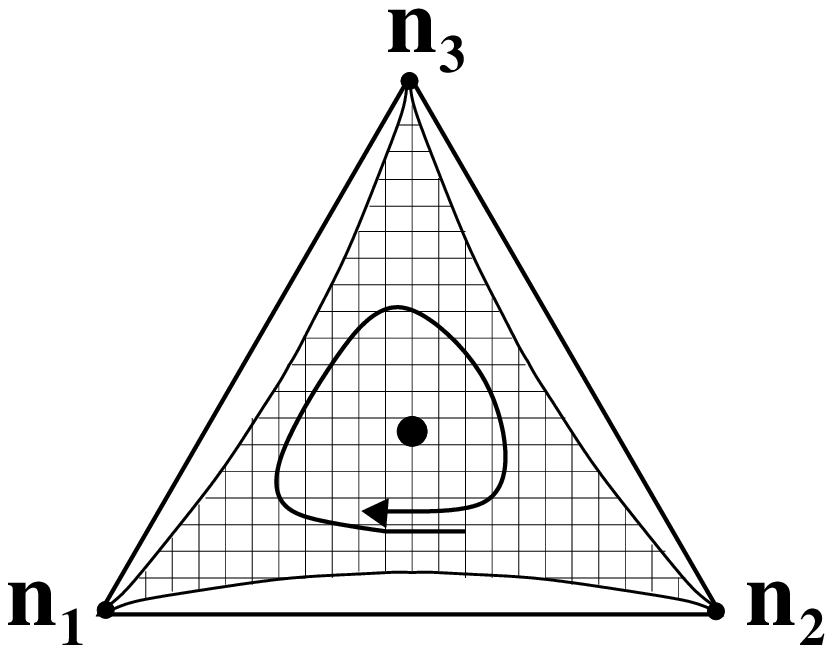,width=6cm}

FIG.\ 7. Dynamic behavior of the cyclic tournament
\end{center}

For stationary points near the middle of FIG.~7 the damping is very small
and cyclic behavior dominates the dynamics. Especially in the middle of
the diagram, i.e.\ for equal rates $r_{1}=r_{2}=r_{3}=r/3$, the motion is
not damped anymore. In this case, the dynamic system Eq.\ (\ref{68}) is
known as a special form of the May-Leonard model \cite{MaLe75} that has
been extensively studied in mathematical biology
\cite{HoSi88,CPC79,SSW79,HoSo94}. Finally, after a relaxation in the plane
$\bar{n} =\sum_{\alpha }\bar{n} _{\alpha }=r$, there exists another
constant of motion $m=\prod_{\alpha } \bar{n}_{\alpha }$ and the dynamic
behavior is characterized by limit cycles around the neutral stationary
point $\left\{ \bar{n}_{\alpha }^{(0)}=r/3\right\} $.

Summarily we have shown that also in stochastic multispecies models of
populations that evolve near the extinction threshold of all colors and
therefore have many absorbing states, the critical properties at the
multicritical point and at all continuous transitions are governed by the
well known Gribov process (Reggeon field theory) exponents (for a previous
simulational result on a two-species system which seems to agree with our
findings see \cite{CDDM91}). In other regions of the phase diagram of course
more complicated critical behavior may arise such as multicritical points
with different scaling exponents \cite{BaBr96}. The models considered here
have many absorbing states (each combination of colors may go extinct
irrespective of the other ones), and therefore are different from models
which were considered by Grinstein et al.\ \cite{GLB89} where it was shown
that multispecies systems with one absorbing state belong to the Gribov
universality class. In addition we have shown that the universal properties
of interspecies correlations and the phase diagram are determined by totally
asymmetric fixed point values of the renormalized interspecies coupling
constants. This eventually leads to a system working cooperatively. It is
interesting that the asymmetry between the species seems to be the condition
for this cooperation near extinction. The model considered here is a simple
but universal model of such a cooperative society and should therefore have
many applications in all fields of natural and even social science.

Competition and extinction is of course a subject much considered in
theoretical bi\-o\-lo\-gy. The main differences between the present work and
the topics covered e.g.\ in the monograph of Hofbauer and Sigmund \cite
{HoSi88} are the more realistic local description of the interactions
between the species, their diffusional motion, and the inclusion of local
fluctuations. Thus here the equations of motion are local stochastic partial
differential equations. Coarse graining and renormalization lead to an
universal macroscopic picture of cooperativity near the critical states of
extinction.

\acknowledgments We thank Uwe T\"auber for many fruitful
discussions, Stephan Theiss for a critical reading of the paper
and Beate Schmittmann for numerous valuable remarks leading to the
final version of the manuscript. This work has been supported in
part by the SFB 237 (``Unordnung und gro\ss e Fluktuationen'') of
the Deutsche Forschungsgemeinschaft.

\appendix

\section{Two-loop Integrals}

In the calculation we will encounter momentum integrals of the type
\begin{equation}
I_{kl;m}=G_{\varepsilon }^{\,-2}\tau ^{\varepsilon }\int\limits_{{\bf
q}_{1}, {\bf q}_{2}}\frac{1}{\Bigl(q_{1}^{\,2}+\tau
\Bigr)^{k}\Bigl(q_{2}^{\,2}+\tau
\Bigr)^{l}\Bigl(q_{1}^{\,2}+q_{2}^{\,2}+(q_{1}+q_{2})^{2}+3\tau
\Bigr)^{m}} \label{A1}
\end{equation}
where $G_{\varepsilon }=\Gamma (1+\varepsilon /2)/(4\pi )^{d/2}$, $
\varepsilon =4-d$, and $\int_{{\bf q}}\ldots =(2\pi )^{-d}\int
d^{d}q\ldots $ . They can be derived from two ``mother'' integrals
\begin{eqnarray}
M^{(1)}(a,b;c) &=&G_{\varepsilon
}^{\,-2}\int\limits_{q_{1},q_{2}}\frac{1}{
\Bigl(q_{1}^{\,2}+a\Bigr)\Bigl(q_{2}^{\,2}+b\Bigr)\Bigl(q_{1}^{\,2}+q_{2}^{
\,2}+(q_{1}+q_{2})^{2}+c\Bigr)}\ ,  \nonumber \\ M^{(2)}(a;c)
&=&G_{\varepsilon }^{\,-2}\int\limits_{q_{1},q_{2}}\frac{1}{
\Bigl(q_{1}^{\,2}+a\Bigr)\Bigl(q_{1}^{\,2}+q_{2}^{\,2}+(q_{1}+q_{2})^{2}+c
\Bigr)}\   \label{A2}
\end{eqnarray}
by taking derivatives with respect to the parameters $a,b,c$. Discarding
nonsingular terms, we find in dimensional regularization
\begin{eqnarray}
M^{(1)}(a,b;c) &=&-\frac{1}{\varepsilon }\Biggl(\frac{a^{1-\varepsilon
}+b^{1-\varepsilon }}{\varepsilon }+\frac{3(a+b)}{2}\biggl(1-\ln
\frac{4}{3} \biggr)+c\ln \frac{4}{3}\Biggr)\ ,  \nonumber \\ M^{(2)}(a;c)
&=&\frac{1}{4\varepsilon }\Biggl(\frac{2c-3a}{\varepsilon }
a^{1-\varepsilon }+ac\biggl(1+\ln
\frac{4}{3}\biggr)-3a^{2}\biggl(1+\frac{1}{ 2}\ln
\frac{4}{3}\biggr)+\frac{c^{2}}{3}\Biggr)\ .  \label{A3}
\end{eqnarray}
These formulas yield the singular parts of the integrals (\ref{A1}) as
\begin{eqnarray}
I_{11;1}^{(SP)} &=&-\frac{1}{\varepsilon ^{2}}\biggl(2+3\varepsilon \biggr)\
,\qquad I_{11;2}^{(SP)}=\frac{1}{\varepsilon }\ln \frac{4}{3}\ ,  \nonumber
\\
I_{12;1}^{(SP)} &=&\frac{1}{2\varepsilon ^{2}}\biggl(2+\varepsilon
-3\varepsilon \ln \frac{4}{3}\biggr)\ ,\qquad I_{10;3}^{(SP)}=\frac{1}{
12\varepsilon }\ ,\qquad I_{20;1}^{(SP)}=\frac{3}{2\varepsilon }\ ,
\nonumber \\ I_{20;2}^{(SP)} &=&\frac{1}{4\varepsilon
^{2}}\biggl(2-\varepsilon +\varepsilon \ln \frac{4}{3}\biggr)\ ,\qquad
I_{30;1}^{(SP)}=-\frac{3}{ 8\varepsilon ^{2}}\biggl(2+\varepsilon
+\varepsilon \ln \frac{4}{3}\biggr)\ . \label{A4}
\end{eqnarray}

\section{Two-loop Selfenergy Diagrams}

\bigskip

\begin{center}
\epsfig{file=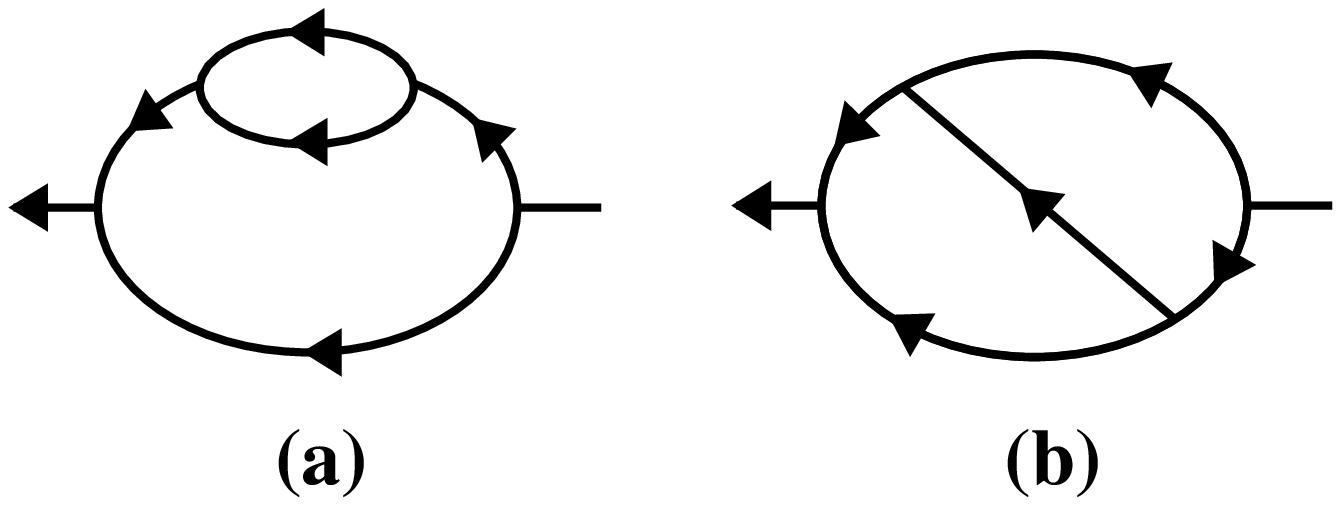,width=8cm}

FIG.\ 8. Two-loop selfenergy diagrams
\end{center}

In FIG.~8 the two-loop selfenergy diagrams are drawn. Diagram FIG.~(8a)
leads to
\begin{eqnarray}
4(a) &=&\frac{(\lambda g)^{4}}{2}\int\limits_{{\bf q}_{1},{\bf q}_{2}}\int
\prod_{i=1}^{3}dt_{i}\text{e}^{-i\omega t_{1}}G({\bf q}/2+{\bf q}
_{1},t_{1})G({\bf q}/2-{\bf q}_{1},t_{1}-t_{2})G({\bf q}/2-{\bf q}_{1},t_{3})
\nonumber \\
&&\hspace{4cm}\times G({\bf q}/4+{\bf q}_{2},t_{1}-t_{3})G({\bf q}/4-{\bf q}
_{1}-{\bf q}_{2},t_{1}-t_{3})\   \label{A5}
\end{eqnarray}
where $G({\bf q},t)$ is the propagator (Eq.\ (\ref{6})) and we always set
$ \tau _{\alpha }=\tau >0$ as an IR regulator. Noting that e$^{-i\omega
t_{1}}G(t_{1})=\bigl($e$^{-i\omega
(t_{1}-t_{2})}G(t_{1}-t_{2})\bigr)\bigl($ e $^{-i\omega
(t_{2}-t_{3})}G(t_{2}-t_{3})\bigr)\bigl($e$^{-i\omega
t_{3}}G(t_{3})\bigr)$ factorizes in the integral, the time integrations
over the intervals $(t_{1}-t_{2})$, $(t_{2}-t_{3})$, $t_{3}$ are easily
performed. The expansion in $i\omega $ and $q^{2}$ to linear order
eventually yields

\begin{eqnarray}
4(a) &=&\frac{\lambda g^{4}}{8}G_{\varepsilon }^{\,2}\tau ^{-\varepsilon }
\biggl(\tau I_{20;1}-\frac{i\omega }{\lambda
}\Bigl(I_{30;1}+I_{20;2}\Bigr) \nonumber \\
&&-\frac{q^{2}}{2}\Bigl(I_{30;1}+\frac{3}{4}I_{20;2}+\frac{1}{2d}\bigl(
I_{20;3}-I_{10;3}\bigr)\Bigr)\biggr)  \label{A6}
\end{eqnarray}
where the $I_{kl;m}$ denote the integrals defined in Eq.\ (\ref{A1}).
Extracting the singular parts using Eq.\ (\ref{A4}), we obtain
\begin{equation}
4(a)=\frac{\lambda g^{4}}{32\varepsilon }G_{\varepsilon }^{\,2}\tau
^{-\varepsilon }\biggl(6\tau +\Bigl(\frac{2}{\varepsilon }+5+\ln
\frac{4}{3} \Bigr)\frac{i\omega }{2\lambda }+\Bigl(\frac{3}{\varepsilon
}+\frac{55}{12}+ \frac{3}{2}\ln \frac{4}{3}\Bigr)\frac{q^{2}}{4}\biggr)\ .
\label{A7}
\end{equation}

In the same way we calculate the second selfenergy diagram FIG.~8(b):
\begin{eqnarray}
4(b) &=&(\lambda g)^{4}\int\limits_{{\bf q}_{1},{\bf q}_{2}}\int
\prod_{i=1}^{3}dt_{i}\text{e}^{-i\omega t_{1}}G({\bf q}/2+{\bf q}
_{1},t_{1}-t_{2})G({\bf q}/2-{\bf q}_{1},t_{1}-t_{3})  \nonumber \\
&&\hspace{2cm}\times G({\bf q}_{1}-{\bf q}_{2},t_{2}-t_{3})G({\bf
q}/2+{\bf q }_{2},t_{2})G({\bf q}/2-{\bf q}_{2},t_{3})  \nonumber \\ &=&\
\frac{\lambda g^{4}}{4}G_{\varepsilon }^{\,2}\tau ^{-\varepsilon }
\biggl(\tau I_{11;1}-\frac{i\omega }{\lambda
}\Bigl(I_{12;1}+I_{11;2}\Bigr) \nonumber \\
&&-\frac{q^{2}}{2}\Bigl(I_{12;1}+I_{11;2}+\frac{2}{d}\bigl(
I_{11;3}+2I_{10;3}-I_{11;2}\bigr)\Bigr)\biggr)  \label{A8}
\end{eqnarray}
up to higher orders in $i\omega $ and $q^{2}$. Extracting the singular parts
again, we find
\begin{equation}
4(b)=-\frac{\lambda g^{4}}{2\varepsilon }G_{\varepsilon }^{\,2}\tau
^{-\varepsilon }\biggl(\Bigl(\frac{1}{\varepsilon }+\frac{3}{2}\Bigr)\tau
+\Bigl(\frac{2}{\varepsilon }+1-\ln \frac{4}{3}\Bigr)\frac{i\omega }{
4\lambda }+\Bigl(\frac{1}{\varepsilon }+\frac{7}{12}-\ln \frac{4}{3}\Bigr)
\frac{q^{2}}{4}\biggr)  \label{A9}
\end{equation}
Summing up, we finally get from $4(a)$ and $(4b)$ the two-loop contribution
of the selfenergy as
\begin{equation}
\Gamma _{\alpha \alpha }^{(2-loop)}=\frac{\lambda g^{4}G_{\varepsilon
}^{\,2} }{\varepsilon ^{2}}\tau ^{-\varepsilon }\biggl(C_{\tau }\tau
+C_{\omega } \frac{i\omega }{\lambda }+C_{q^{2}}q^{2}\biggr)\ ,  \label{A9a}
\end{equation}
where
\begin{equation}
C_{\tau }=\frac{1}{2}+\frac{9\varepsilon }{16}\ ,\quad C_{\omega }=\frac{7}{
32}\Bigl(1+\frac{3\varepsilon }{14}-\frac{9\varepsilon }{14}\ln \frac{4}{3}
\Bigr)\ ,\quad C_{q^{2}}=\frac{13}{128}\Bigl(1+\frac{19\varepsilon }{52}-
\frac{35\varepsilon }{26}\ln \frac{4}{3}\Bigr)\ .  \label{A9b}
\end{equation}

\section{Two-loop Vertex Diagrams}

\bigskip

\bigskip

\begin{center}
\epsfig{file=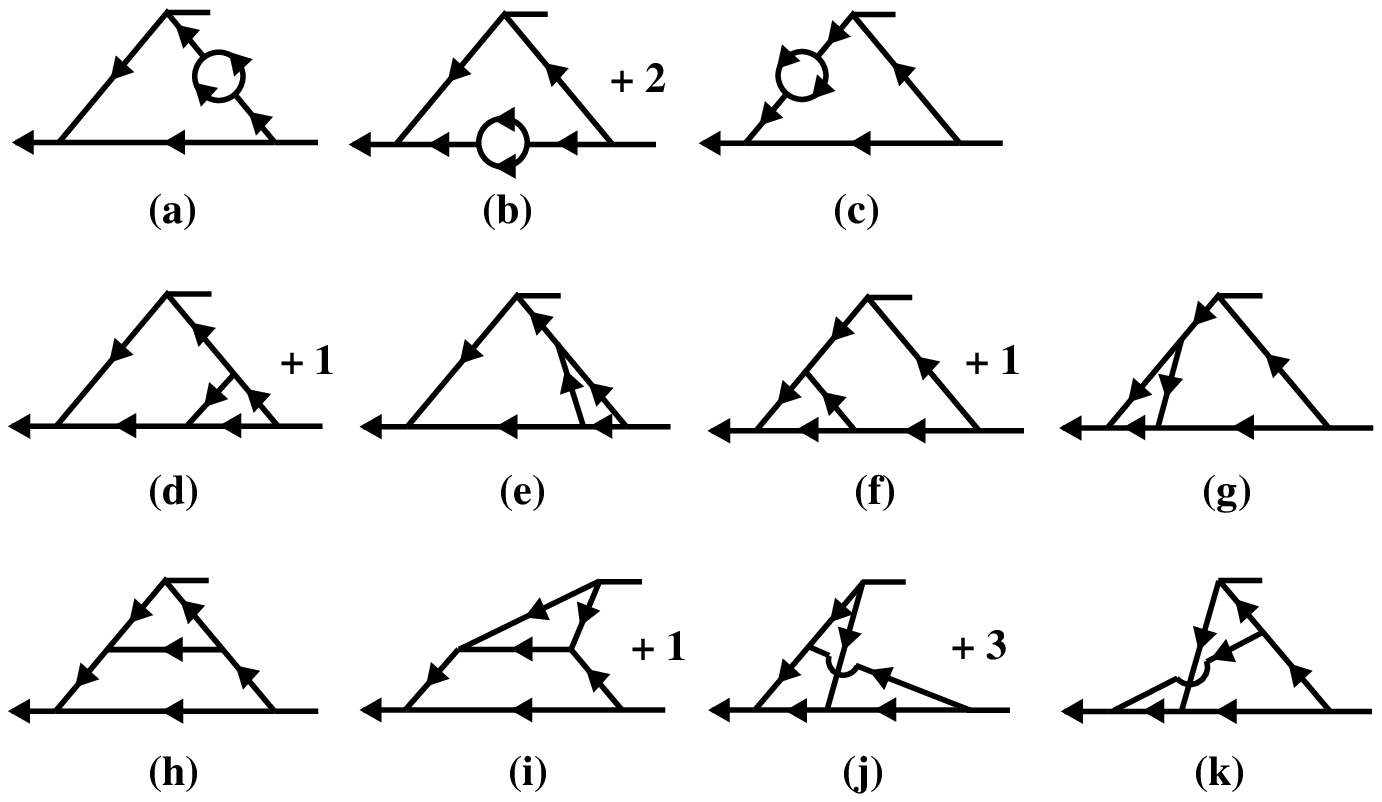,width=12cm}

FIG.\ 9. Two-loop vertex diagrams; the numbers show the other time
orderings
\end{center}

FIG.~9 presents the eleven two-loop vertex diagrams. They can be
calculated by the same method as the selfenergy diagrams. Here the
external frequencies and momenta can be set to zero because the expansion
in these variables does not lead to primitive divergencies. As an example
we show the calculation of the diagram FIG.~9(i) explicitly.

The two possible different time orderings of the vertices are shown in
FIG.~10.

\bigskip

\begin{center}
\epsfig{file=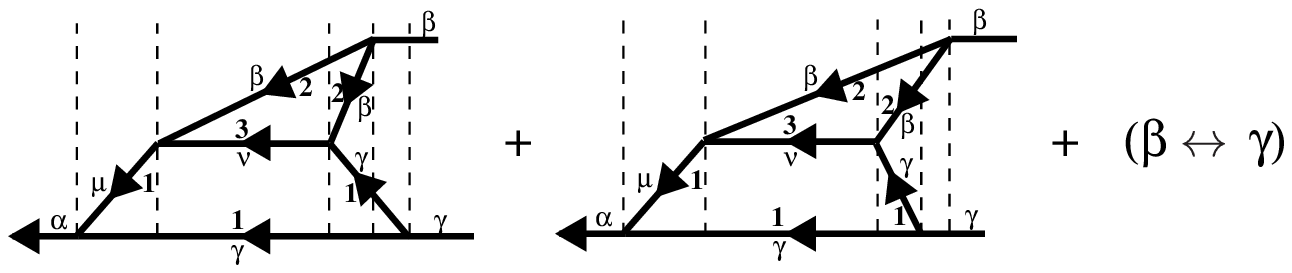,width=12cm}

FIG.\ 10. Two different time orderings of a vertex diagram
\end{center}
The symmetry factor of the diagrams is one, but one also has to add the
two diagrams arising from the interchange of the color indices $\beta $
and $ \gamma $. After the factorization of the propagators, the
integration over the time intervals, indicated by the broken lines, is
trivial und leads to the expression (with the abbreviations $\kappa
_{i}=\lambda (\tau +q_{i}^{\,2})$, ${\bf q}_{3}={\bf q}_{1}+{\bf q}_{2}$)
\begin{eqnarray}
5(i) &=&-\frac{\lambda ^{5}}{8}\sum_{\mu ,\nu }\bigl(\delta _{\alpha \mu
}g_{\alpha \nu }+\delta _{\alpha \gamma }g_{\alpha \mu }\bigr)\bigl(\delta
_{\mu \nu }g_{\mu \beta }+\delta _{\mu \beta }g_{\mu \nu
}\bigr)\bigl(\delta _{\alpha \mu }g_{\alpha \nu }+\delta _{\alpha \gamma
}g_{\alpha \mu }\bigr) g_{\beta }g_{\gamma }  \nonumber \\ &&\times
\int\limits_{{\bf q}_{1},{\bf q}_{2}}\biggl(\frac{1}{(2\kappa _{1})(\kappa
_{1}+\kappa _{2}+\kappa _{3})(2\kappa _{1}+2\kappa _{2})(2\kappa _{1})}
\nonumber \\ &&\hspace{1.5cm}+\frac{1}{(2\kappa _{1})(\kappa _{1}+\kappa
_{2}+\kappa _{3})(2\kappa _{1}+2\kappa _{2})(2\kappa _{2})}\biggr)+(\beta
\leftrightarrow \gamma )  \nonumber \\ &=&-\frac{\lambda
g^{2}}{32}G_{\varepsilon }^{\,2}\tau ^{-\varepsilon } \bigl( \delta
_{\alpha \beta }g_{\alpha \gamma }+\delta _{\alpha \gamma }g_{\alpha \beta
}\bigr)\Bigl(g\bigl(g_{\beta \gamma }+g_{\gamma \beta } \bigr) +g_{\alpha
\beta }g_{\alpha \gamma }+g_{\beta \gamma }g_{\gamma \beta
}\Bigr)I_{12,1}\ .  \label{A10}
\end{eqnarray}
Thus we obtain the contribution of the diagram FIG.~9(i) to the vertex
function $\Gamma _{\alpha ,\alpha \beta }$:
\begin{equation}
\Gamma _{\alpha ,\alpha \beta }^{5(i)}=\frac{\lambda g^{2}}{16}
G_{\varepsilon }^{\,2}\tau ^{-\varepsilon }g_{\alpha \beta }\Bigl(g\bigl(
g_{\beta \alpha }+g_{\alpha \beta }\bigr)+g_{\alpha \beta }g+g_{\beta
\alpha }g_{\alpha \beta }\Bigr)I_{12,1}\ .  \label{A11}
\end{equation}

In the same manner we calculate the other diagrams of FIG.~9 and find
\begin{eqnarray}
\Gamma _{\alpha ,\alpha \beta }^{5(a)}&=&\frac{\lambda g^{2}}{32}
G_{\varepsilon }^{\,2}\tau ^{-\varepsilon }g_{\alpha \beta
}g\Bigl(2g+g_{\alpha \beta }+g_{\beta \alpha }\Bigr)I_{30,1\ ,}  \label{A12}
\\
\Gamma _{\alpha ,\alpha \beta }^{5(b)}&=&\frac{\lambda g^{2}}{16}
G_{\varepsilon }^{\,2}\tau ^{-\varepsilon }g_{\alpha \beta
}g\Bigl(2g+g_{\alpha \beta }+g_{\beta \alpha
}\Bigr)\Bigl(I_{20,2}+I_{30,1}\Bigr)\ ,  \label{A13} \\
\Gamma _{\alpha ,\alpha \beta }^{5(c)}&=&\frac{\lambda g^{2}}{32}
G_{\varepsilon }^{\,2}\tau ^{-\varepsilon }g_{\alpha \beta
}g\Bigl(2g+g_{\alpha \beta }+g_{\beta \alpha }\Bigr)I_{30,1}\ ,  \label{A14}
\\
\Gamma _{\alpha ,\alpha \beta }^{5(d)}&=&\frac{\lambda g^{2}}{16}
G_{\varepsilon }^{\,2}\tau ^{-\varepsilon }g_{\alpha \beta
}g\Bigl(2g+g_{\alpha \beta }+g_{\beta \alpha
}\Bigr)\Bigl(2I_{11,2}+I_{12,1}\Bigr)\ ,  \label{A15} \\
\Gamma _{\alpha ,\alpha \beta }^{5(e)}&=&\frac{\lambda g^{2}}{16}
G_{\varepsilon }^{\,2}\tau ^{-\varepsilon }g_{\alpha \beta
}g\Bigl(2g+g_{\alpha \beta }+g_{\beta \alpha }\Bigr)I_{12,1}\ ,  \label{A16}
\\
\Gamma _{\alpha ,\alpha \beta }^{5(f)}&=&\frac{\lambda g^{2}}{32}
G_{\varepsilon }^{\,2}\tau ^{-\varepsilon }g_{\alpha \beta }\Bigl(\bigl(
g_{\alpha \beta }^{\,2}+g_{\beta \alpha }^{\,2}\bigr)+2gg_{\beta \alpha
}+4g^{\,2}\Bigr)\Bigl(2I_{11,2}+I_{12,1}\Bigr)\ ,  \label{A17} \\ \Gamma
_{\alpha ,\alpha \beta }^{5(g)}&=&\frac{\lambda g^{2}}{16} G_{\varepsilon
}^{\,2}\tau ^{-\varepsilon }g_{\alpha \beta }\Bigl(g\bigl( 2g+g_{\alpha
\beta }\bigr)+g_{\alpha \beta }g_{\beta \alpha }\Bigr)I_{12,1}\ .
\label{A18} \\ \Gamma _{\alpha ,\alpha \beta }^{5(h)}&=&\frac{\lambda
g^{2}}{16} G_{\varepsilon }^{\,2}\tau ^{-\varepsilon }g_{\alpha \beta
}\Bigl(g_{\alpha \beta }^{\,2}+g_{\beta \alpha }^{\,2}+2gg_{\beta \alpha
}+4g^{\,2}\Bigr)I_{20,2}\ ,  \label{A19} \\ \Gamma _{\alpha ,\alpha \beta
}^{5(j)}&=&\frac{3\lambda g^{2}}{16} G_{\varepsilon }^{\,2}\tau
^{-\varepsilon }g_{\alpha \beta }^{\,2}\Bigl(g+g_{\beta \alpha
}\Bigr)I_{11,2}\ ,  \label{A20} \\ \Gamma _{\alpha ,\alpha \beta
}^{5(k)}&=&\frac{\lambda g^{2}}{16} G_{\varepsilon }^{\,2}\tau
^{-\varepsilon }g_{\alpha \beta }\Bigl(g_{\alpha \beta }^{\,2}+g_{\beta
\alpha }^{\,2}+2gg_{\beta \alpha }+4g^{\,2}\Bigr)I_{11,2}\ .  \label{A21}
\end{eqnarray}

Adding up all the two-loop contributions $\Gamma _{\alpha ,\alpha \beta
}^{5(a)},\cdots \Gamma _{\alpha ,\alpha \beta }^{5(k)}$ and extracting the
singular parts Eq.\ (\ref{A4}), we finally get
\begin{eqnarray}
\Gamma _{\alpha ,\alpha \beta }^{(2-loop)} &=&\frac{\lambda
g^{2}G_{\varepsilon }^{\,2}}{16\varepsilon }\tau ^{-\varepsilon }g_{\alpha
\beta }\biggl(\Bigl(\frac{8}{\varepsilon }+1\Bigr)g^{2}+\Bigl(\frac{4}{
\varepsilon }+\frac{1}{2}\Bigr)gg_{\beta \alpha
}+\Bigl(\frac{2}{\varepsilon }+1\Bigr)g_{\alpha \beta }g_{\beta \alpha }
\nonumber \\ &&+\Bigl(\frac{4}{\varepsilon }+\frac{3}{2}-3\ln
\frac{4}{3}\Bigr)gg_{\alpha \beta }+\Bigl(\frac{1}{\varepsilon
}+\frac{3}{2}\ln \frac{4}{3} \Bigr)\Bigl(g_{\alpha \beta }^{\,2}+g_{\beta
\alpha }^{\,2}\Bigr)\biggr) +O(\varepsilon ^{0})\ .  \label{A22}
\end{eqnarray}











\end{document}